\documentclass[11pt,a4paper]{article}
\usepackage[latin1]{inputenc}
\usepackage[T1]{fontenc}
\usepackage{amsfonts}
\usepackage{amssymb}
\usepackage[left=25mm,top=25mm,bottom=25mm,right=25mm]{geometry}
\usepackage{dsfont,amscd}
\usepackage{graphicx}
\usepackage[cyr]{aeguill}
\usepackage[pdfdisplaydoctitle=true,
            colorlinks=true,
            urlcolor=blue,
            citecolor=blue,
            linkcolor=blue,
            pdfstartview=FitH,
            pdfpagemode=None,
            bookmarksnumbered=true]{hyperref}
\usepackage{amssymb}
\usepackage{amsmath}
\usepackage{todonotes}
\usepackage{amsthm}
\usepackage{float}
\usepackage{arydshln}
\usepackage{subfigure}
\usepackage[all]{xy}
\graphicspath{{fig/}}
\theoremstyle{definition}

\theoremstyle{remark}

\newcommand{\SO}{\mathrm{SO}}
\newcommand{\OO}{\mathrm{O}}

\newcommand{\ZZ}{\mathrm{Z}}
\newcommand{\DD}{\mathrm{D}}

\newcommand{\Id}{\mathrm{Id}}

\newcommand{\V}[1]{\underline{\mathrm{#1}}}
\newcommand{\dT}[1]{\underset{\sim}{\mathrm{#1}}}
\newcommand{\tTd}[1]{\underset{\simeq}{\mathrm{#1}}}

\newcommand{\qT}[1]{\underset{\approx}{\mathrm{#1}}}
\newcommand{\cTd}[1]{\underset{\approxeq}{\mathrm{#1}}}
\newcommand{\sT}[1]{\underset{\underset{\sim}{\approx}}{\mathrm{#1}}}

\newcommand{\be}{\begin{equation}}
\newcommand{\ee}{\end{equation}}
\newcommand{\ben}{\begin{equation*}}
\newcommand{\een}{\end{equation*}}
\newcommand{\ba}{\begin{eqnarray}}
\newcommand{\ea}{\end{eqnarray}}
\newcommand{\ban}{\begin{eqnarray*}}
\newcommand{\ean}{\end{eqnarray*}}

\usepackage{authblk}

\title{Anisotropic and dispersive wave propagation within strain-gradient framework}

\author[a]{G. Rosi}

\author[b]{N. Auffray}
\date{}
\affil[a]{Universit\'e Paris-Est, Laboratoire Mod\'elisation et Simulation Multi Echelle, 
MSME UMR 8208 CNRS, 61 av du G\'en\'eral de Gaulle, 94010 Cr\'eteil Cedex, France}

\affil[b]{Universit\'e Paris-Est, Laboratoire Mod\'elisation et Simulation Multi Echelle, 
MSME UMR 8208 CNRS, 5 bd Descartes, 77454 Marne-la-Vall\'ee, France}
\begin{document}
\maketitle

\begin{abstract}
In this paper anisotropic and dispersive wave propagation within linear strain-gradient elasticity is investigated. This analysis reveals significant features of this extended theory of continuum elasticity. First, and contrarily to classical elasticity, wave propagation in hexagonal (chiral or achiral) lattices becomes anisotropic as the frequency increases. Second, since strain-gradient elasticity is dispersive, group and energy velocities have to be treated as different quantities. These points are first theoretically derived, and then numerically experienced on hexagonal chiral and achiral lattices. The use of a continuum model for the description of the high frequency behavior of these microstructured materials can be of great interest in engineering applications, allowing problems with complex geometries to be more easily treated. 
\end{abstract}

\noindent \textbf{Keywords:} Strain gradient elasticity , Anisotropy , Higher-order tensors , Chirality , Acoustical activity , Wave propagation

\section{Introduction}

The study of wave propagation within Periodic Architectured Materials (PAM) is a topic of increasing interest. This subject finds its origin in the field of electromagnetism, where it drove the development of innovative materials and devices, e.~g. smart wave guides or cloaking devices \cite{SMJ+06}. Indeed, materials with exotic properties (e.~g.  stop bands, energy focusing) are obtained by exploiting the periodic nature of such materials. 
The same concept can be successfully applied to elastic waves for designing materials capable of changing the direction of propagation of the energy (e.~g. wave beaming \cite{RSS03}), to enhance the non-destructive characterization properties of the material itself (e.~g. materials with a specific acoustic signature when damaged \cite{Madeo:2014ge}.

\begin{figure}[ht]
\centering
\subfigure[Low frequency (800 Hz)]{\includegraphics[width=0.4\linewidth]{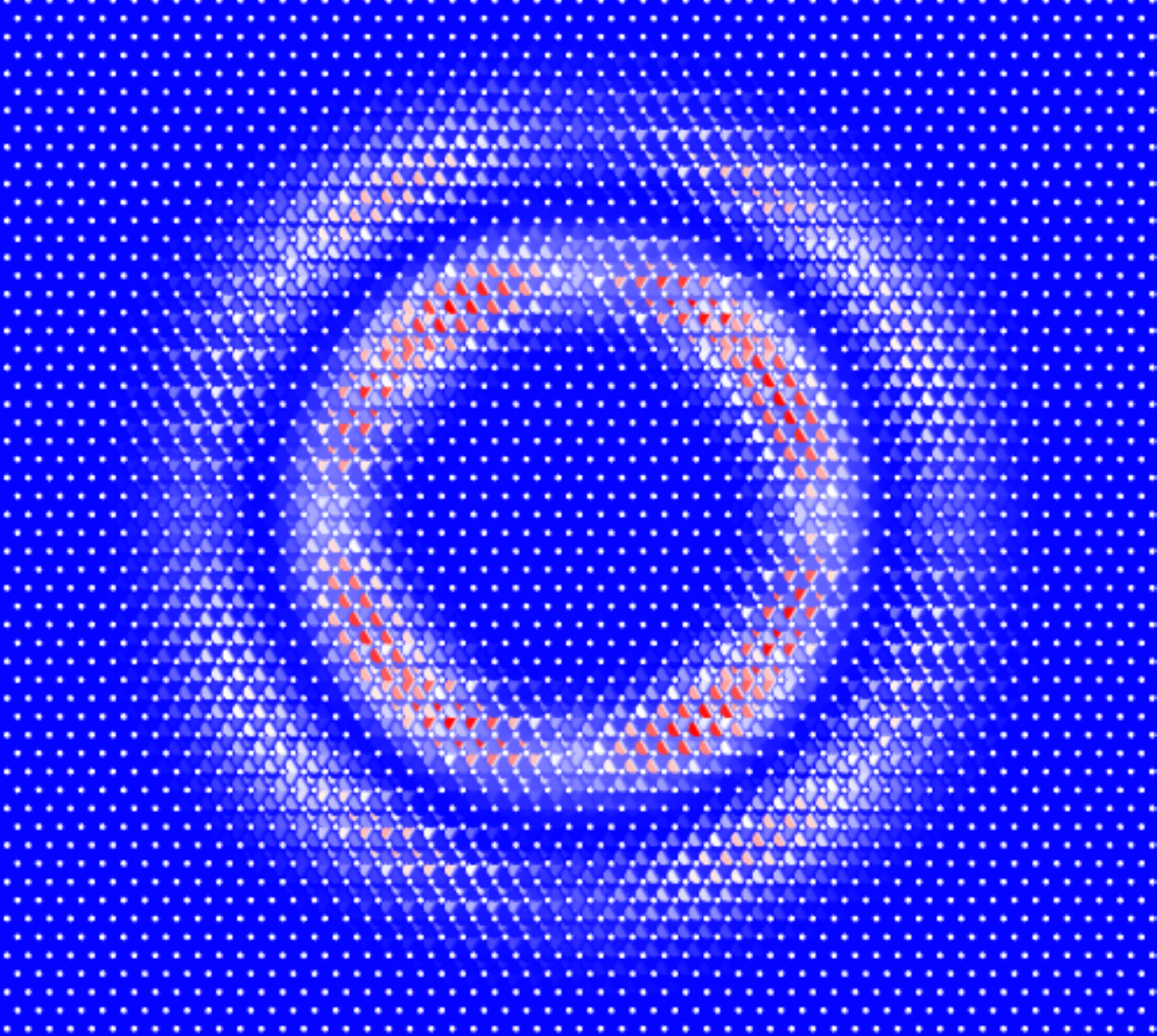}\label{fig:Snapshots-long}}\qquad
\subfigure[High frequency (2 kHz)]{\includegraphics[width=0.4\linewidth]{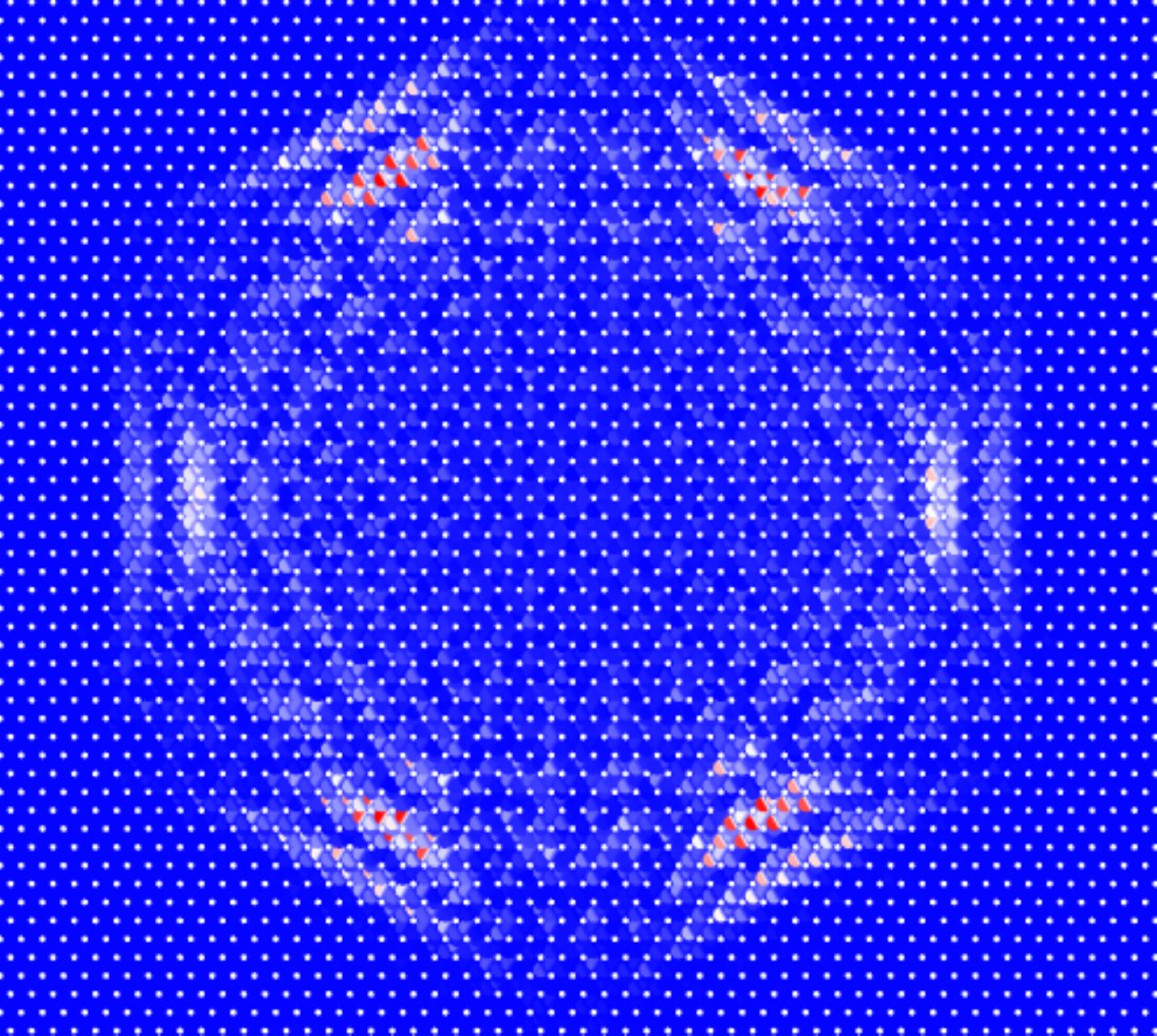}\label{fig:Snapshots-short}}
\caption{Time transient Finite Elements simulation of energy propagation in a hexagonal lattice at a given instant for shear pulses of different central frequency (color online).}
\centering
\label{fig:Snapshots}
\end{figure}

The behavior of waves propagating in these media strongly depends on frequency. For example some of them have an isotropic behavior at low frequencies and become anisotropic at high frequencies. Well known is the case of hexagonal lattices, used in so-called honeycomb structures, for which an isotropic (in 2D) or a transverse isotropic (in 3D) model is commonly used. Indeed, when performing a simple wave propagation test, a breaking of symmetry  occurs when frequency increases. To illustrate this phenomenon, we use the results of a Finite Elements simulation performed on the full honeycomb geometry modeled by clamped Timoshenko beams. We observe the evolution of the total energy when shear pulses of different central frequencies are applied at the center of the structure. Two snapshots at suitably chosen time instants are plotted in Fig.\ref{fig:Snapshots}. In the case of low frequencies (we chose the value of 800 Hz, to avoid boundary effects for longer wavelenghts) the propagation is isotropic (Fig.\ref{fig:Snapshots-long}). However, when increasing the frequency up to 2 KHz, the breaking of symmetry occurs, revealing the inherent symmetries of the hexagonal lattice (Fig.\ref{fig:Snapshots-short}). We can also observe the effects of energy focusing at discrete orientations corresponding to pure modes of propagation \cite{Wol05}. This phenomenon has also been experimentally observed (e.~g. in \cite{CG14}). In a perspective of a homogenization procedure, this behavior should be reproduced by any Homogeneous Equivalent Medium (HEM).

In its basic formulation, elastic wave propagation within PAM shares many aspects with lattice dynamics. Indeed, tools classically used in condensed matter physics can be also employed for the study of such a class of materials \cite{DDJ08,GDK+13,NHA15a}. For example, dispersion curves of PAMs are usually obtained by complete  computations, following the Bloch theorem \cite{PWF06,SRG+09,LHS+11,LHH12}, on the unit cell. 

\begin{figure}[ht]
\centering
\subfigure[Hexachiral lattice]{\includegraphics[width=0.3\linewidth]{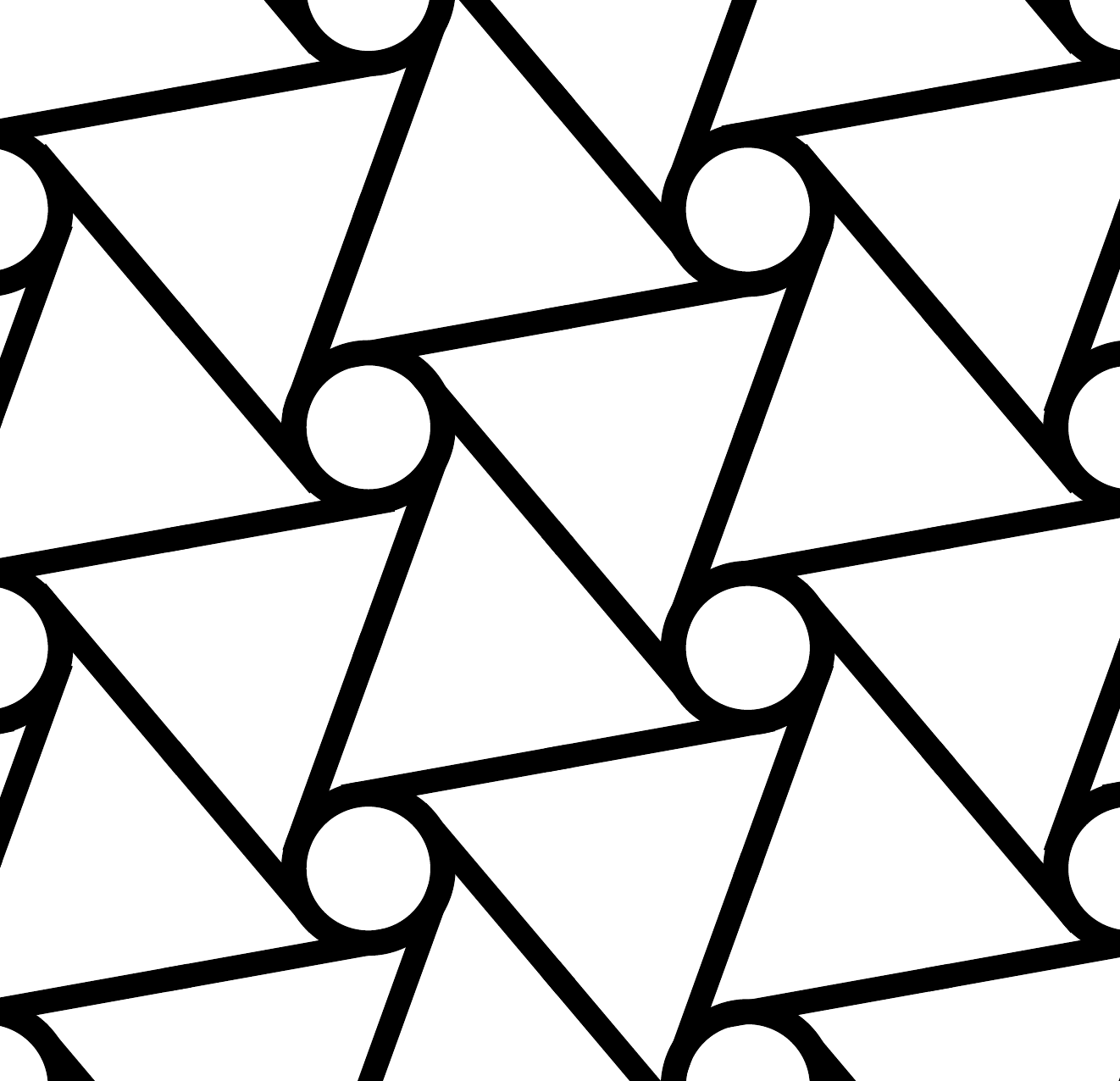}\label{fig:AuxZ6}}\qquad
\subfigure[Dispersion analysis on a hexachiral material]{\includegraphics[width=0.5\linewidth]{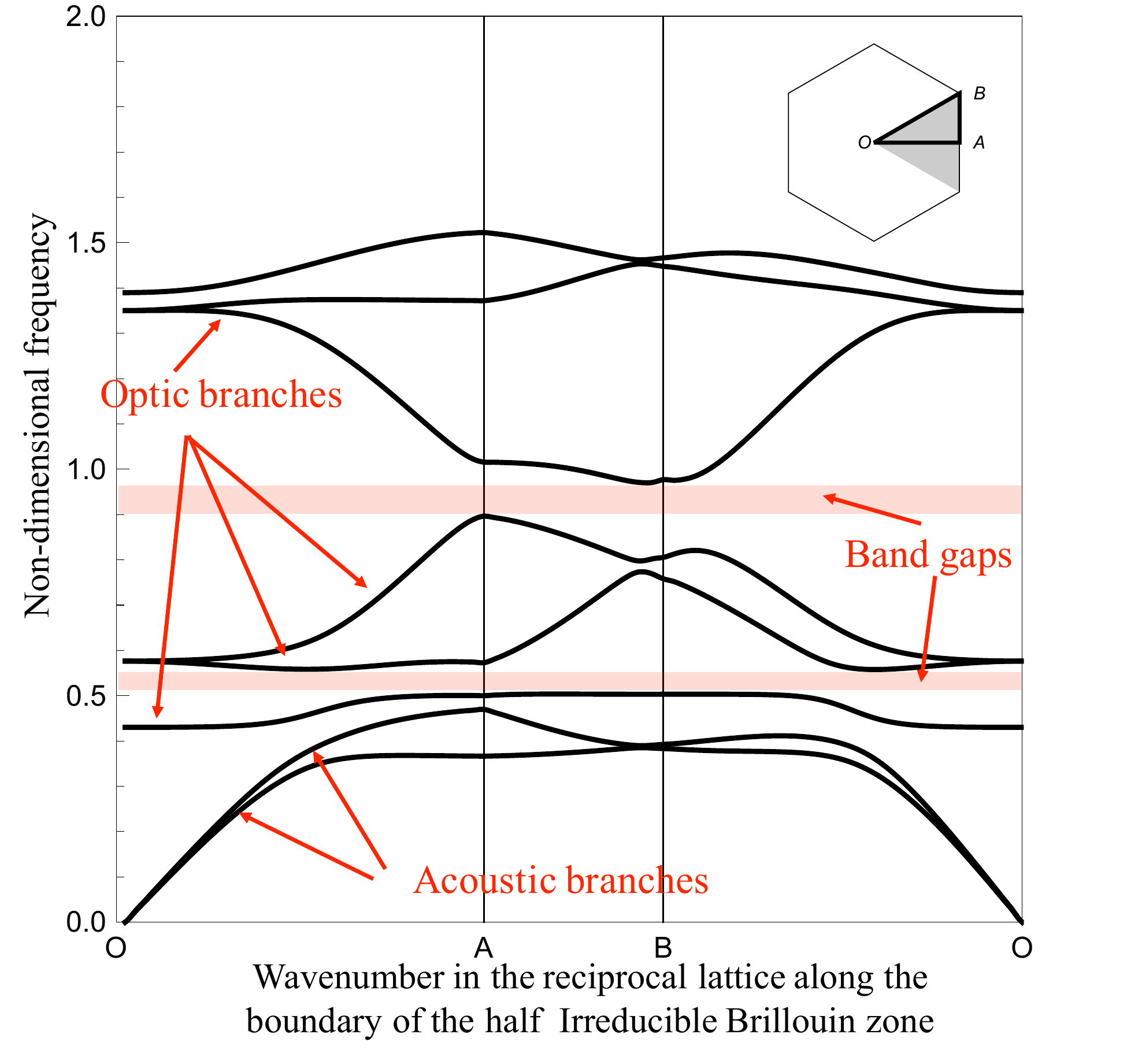}\label{fig:ExaDisp}}
\caption{Geometry and dispersion curves for a hexachiral material.}
\centering
\label{fig:Z6}
\end{figure}

These dispersion curves are of prime importance for the conception of tailored meta-materials, wave guides or sensors \cite{LSS15}, since they allow to link the geometric properties of the unit cell with the dynamic properties of the lattice. For understanding the richness of the response of such materials, it is useful to consider an example. Among the multiple possible choices we chose the hexachiral cell, depicted in Fig. \ref{fig:AuxZ6}. The dispersion diagram obtained by a Finite Element (FE) computation  is presented in Fig. \ref{fig:ExaDisp}. This plot represents the dispersion relation computed on the edges of the irreducible Brillouin (IBZ) zone delimited by the points O, A and B. It captures the essential propagation properties of the material. The behavior is rich, and some key features can be highlighted: i) the presence of acoustic branches, i.e. those starting from the origin; ii) the presence of optic branches, i.e. those exhibiting a cut-off frequency; iii) the presence of frequency band gaps, or stop bands, where no wave can propagate (the wavenumber $k$ is complex); iv) the presence of dispersive behavior. Moreover, since the graph is not perfectly symmetric, i.~e. the path O-A-B-O is not the same as the path O-B-A-O, the material is also experiencing directivity. This means that propagation constants will depend on the direction of propagation.  All these phenomena appear only when increasing the frequency or reducing the wavelength, and may be of crucial importance according to the sought application.

\begin{figure}
\centering
\includegraphics[scale=0.5]{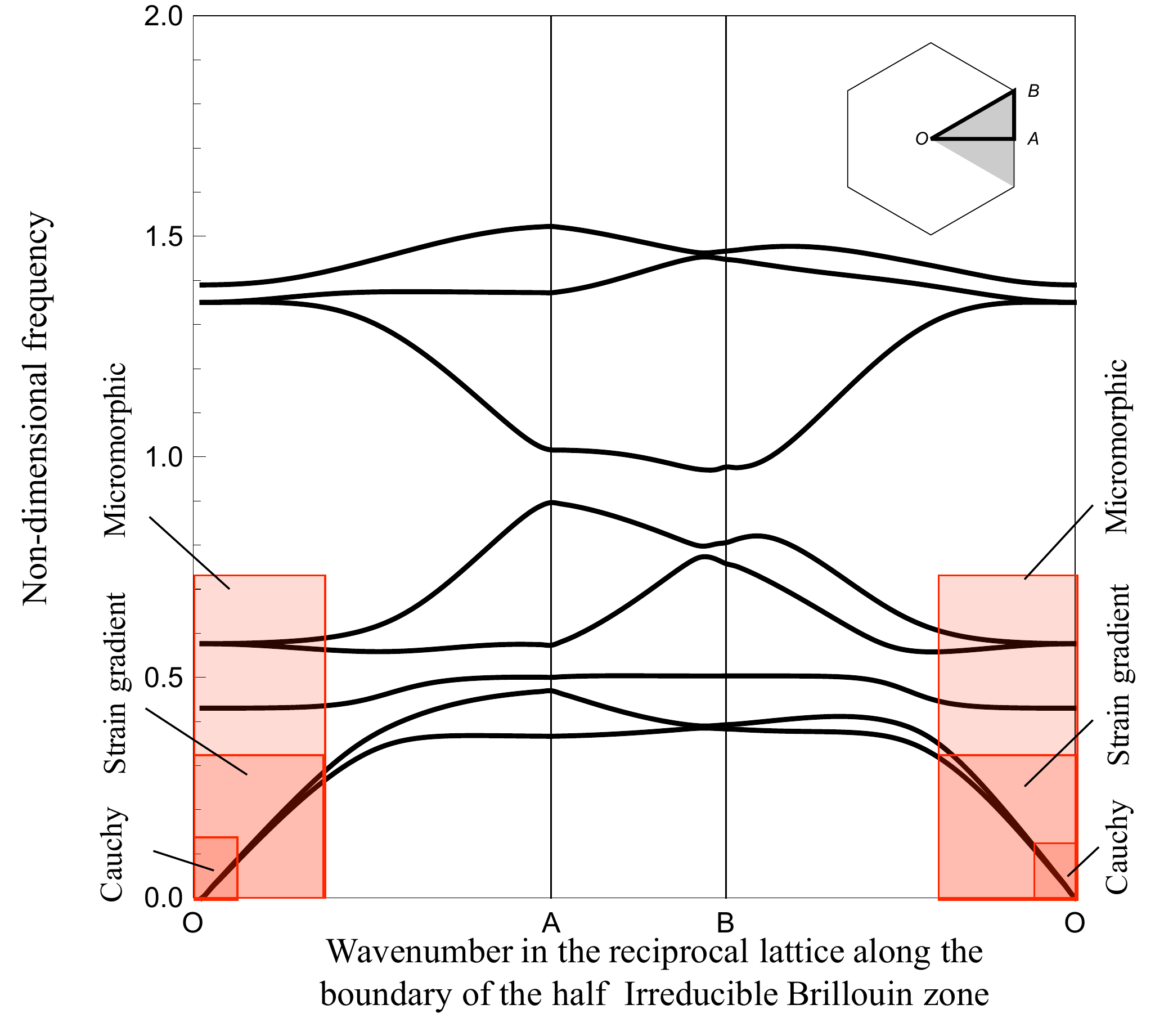}
\caption{Dispersion analysis on a hexachiral material with superposed domains of validity of the continuum models.}\label{fig:HexDispZones}
\end{figure}

For practical applications, e.~g. for simplifying the study of reflection/transmission problems \cite{dellIsola:2011ti,Rosi:2014hk,Rosi:2014il,Gourgiotis:2013gt}, it is of interest to determine a HEM as an approximation of the PAM, provided that some of the specific aforementioned features are preserved.
Classical Cauchy elasticity is a theory of the linear approximation of the acoustic branches. This theory is usually sufficiently accurate to describe the behavior of a homogeneous or slightly heterogeneous anisotropic medium. However, when increasing the frequency or wavenumber, this theory fails to capture almost all the key features of PAMs, as highlighted in the example on hexagonal lattice. In Figure \ref{fig:HexDispZones} the validity zone of Cauchy model is approximately highlighted. A HEM that completely reproduces the dispersion curve implies to use a highly non-local continuum, as the one introduced by Willis \cite{Wil85,Wil97,NHA15a}. But since its use is almost as complex and challenging as the complete problem, local approximations are, in practice, preferred \cite{NHA15b}. 

Local extensions of the classical continuum mechanics are basically of two types\footnote{Those approaches can perfectly be combined.} \cite{Tou62,Min64,Min65,Eri68,ME68}:
\begin{description}
\item[Higher-order continua:] the number of degrees of freedom is extended, and hence optical branches can be modeled. The Cosserat model (also known as micropolar), in which local rotations are added as degrees of freedom, belongs to this family \cite{Cos09}. This enhancement can be extended further to obtain the micromorphic elasticity \cite{GR64,Min64,Ger73,Eri68}.
\item[Higher-grade continua:] the degrees of freedom are kept identical but higher-order gradients of the displacement field are involved into the elastic energy. Within this framework no optical branch is present. Mindlin's  Strain-Gradient Elasticity (SGE) model \cite{Min64,ME68,Min65} belongs to this family.
\end{description}
Strain gradient elasticity can be retreived as a Low Frequency (LF), Long Wave-length (LW) approximation of the micromorphic kinematic \cite{Min64}. As a consequence, the parameters needed to set up this model are limited compared to a complete micromorphic continuum. The domain of validity of these extended theories are roughly estimated in Fig.\ref{fig:HexDispZones}. In the case of micromorphic continua, band gaps can be modelled only when considering the relaxed formulation presented in \cite{Neff:2014ek,Ghiba:2014fo,Madeo:2014df,Madeo:2013uq}.
As can be observed in Fig. \ref{fig:HexDispZones}, in the LF limit all the internal degrees of freedom  are lost, as well as all optical branches.  In Long Wavelength (LW) limit the dispersion relation becomes linear, and hence dispersive effects are lost.  

In the present paper, attention will be devoted to the modeling of anisotropic dispersive elastic waves in the framework of strain-gradient elasticity. This approach can be seen as a phenomenological reformulation of some pioneering works in physics of dispersive elasticity \cite{PB68,DiV86}. This work follows some previous contributions, since the different anisotropic elasticity tensors involved in SGE have been studied in \cite{ABB09,ADR15}. As will be shown, wave propagation within linear SGE allows to capture some specific features that can not be modelled classically:
\begin{enumerate}
\item Chiral sensitivity;
\item Anisotropy of hexagonal lattices;
\item Distinction between group and energy velocities;
\end{enumerate}
It has to be noted that the second point, cannot be modeled using a Cosserat (or Micropolar) medium. All of these effects are controlled by the circular frequency $\omega$ and disappear, as it should, in the LF limit, where classical Cauchy behavior is retrieved. These specificities will be illustrated both analytically and numerically, and the results from the generalized continuum model will be compared with full field simulations.\\

\noindent\textbf{Organisation of the paper}\\ The paper is organized as follows. In a first time, \S.\ref{S:MSGE}, the basic equations of Strain Gradient Elasticity are recapped. In \S.\ref{S:PlW} plane wave propagation in SGE is discussed and the generalized acoustic tensor is introduced. Then, in  \S.\ref{S:HexMat} numerical studies are conducted on hexagonal and hexachiral lattices in order to illustrate these different aspects. Finally, \S.\ref{S:concl} is devoted to conclusions.\\

\noindent\textbf{Notations}\\ In this work  tensors of order ranking from $0$ to $6$ are denoted, respectively, by $\mathrm{a}$, $\V{a}$, $\dT{a}$, $\tTd{a}$, $\qT{a}$, $\cTd{a}$ and $\sT{a}$. The simple, double  and fourth contractions are written $\, ., \, :$ and $::$ respectively. In index form with respect to an orthonormal Cartesian basis,
these notations correspond to
\ben
\V{a} . \V{b} = a_i b_i, \quad \dT{a} \, : \, \dT{b} = a_{ij} b_{ij}, \quad \qT{a} \, :: \, \qT{b} = a_{ijkl} b_{ijkl}
\een
where repeated indices are summed up. Spatial gradient will classically be denoted, in index form, by a comma 
\ben
\mathrm{Grad}\ \V{a}=\left(\V{a}\otimes\V{\nabla}\right)_{ij}=a_{i,j}
\een
Divergence and curl will be denoted:
\ben
\mathrm{Div}\ \V{a}=\left(\V{\nabla}\cdot\V{a}\right)=a_{i,i}\quad ;\quad \mathrm{Curl}\ \V{a}=\left(\V{\nabla}\times\V{a}\right)_{i}=\epsilon_{ijk}a_{j,k}
\een
where $\epsilon_{ijk}$ is the Levi-Civita symbol.
Vector spaces will be denoted using blackboard bold fonts, and their tensorial order indicated by using formal indices. When needed index symmetries are expressed as follows:  $(..)$  indicates invariance under permutations of the indices in parentheses, while $\underline{..}\ \underline{..}$ denotes invariance with respect to permutations of the underlined blocks.   Finally, a superimposed dot will denote a partial time derivative.

\section{Strain-gradient dynamics} \label{S:MSGE}

In this section the Strain Gradient Elasticity (SGE) model will first be recapped, then particular attention will be devoted to the constitutive laws.
\subsection{Dynamics}
In this section  the dynamic balance equations for a strain-gradient solid will be summed-up. Our setting will be based on Mindlin type II formulation \cite{ME68}. Strain-gradient elasticity can be considered as a long wave approximation of the micromorphic model (see \cite{Min64} for more details). The degrees of freedom of the model are the components of the displacement field $u_i$. Kinetic and potential energy densities, respectively denoted $\mathcal{K}$ and $\mathcal{P}$, are  functions of the displacement and its gradients up to the second order:
\ben
\mathcal{K}=\frac{1}{2}p_i v_i+\frac{1}{2}q_{ij} v_{i,j},\qquad
\mathcal{P}=\frac{1}{2}\sigma_{ij}\varepsilon_{ij}+\frac{1}{2}\tau_{ijk}\eta_{ijk},
\een
where 
\begin{itemize}
\item $p_{i}$ and $q_{ij}$ are, respectively, the momentum and the hypermomentum tensors;
\item $v_{i}$ and $v_{i,j}$ are the velocity ($v_{i}=\dot{u}_{i}$) and its gradient;
\item $\sigma_{ij}$ and $\tau_{ijk}$ are the stress and the hyperstress tensors;
\item $\varepsilon_{ij}$ and $\eta_{ijk}=\varepsilon_{ij,k}$, the infinitesimal strain tensor ($\varepsilon_{ij}=(u_{i,j}+u_{j,i})/2$) and its gradient.
\end{itemize}

By application of the least action principle on the action functional \cite{Min64}, and neglecting body double forces, the following bulk equations are obtained
\begin{equation}
s_{ij,j}=\dot{p}_i-\dot{q}_{ij,j}\label{eq:Bulk}
\end{equation}
where $s_{ij}$ is the effective second order symmetric stress tensor, defined as follows:
\be
s_{ij}=\sigma_{ij}-\tau_{ijk,k} \label{eq:effectivestress}
\ee

Consistent boundary conditions are also obtained from the least action principle, but as the objective of this paper is to study free propagation we omit to list them here (for details see \cite{Min64,AIE+13}). To be well posed those equations have to supplemented by a constitutive law.

\subsection{Constitutive law}

In a spatio-temporal framework, the constitutive law reads 
\begin{align}
 & p_{i}=\rho\delta_{ip}v_p+K_{ipq}v_{p,q}\label{eq:Const3}\\
 & q_{ij}=K_{ijp}v_p+J_{ijpq}v_{p,q}\label{eq:Const4}\\
 & \sigma_{ij}=C_{ijpq}\varepsilon_{pq}+M_{ijpqr}\eta_{pqr}\label{eq:Const1}\\
 & \tau_{ijk}=M_{pqijk}\varepsilon_{pq}+A_{ijkpqr}\eta_{pqr}\label{eq:Const2}
\end{align}
or, using a quadratic form
\ben
\begin{pmatrix}
\V{p}\\
\dT{q}\\
\dT{\sigma}\\
\tTd{\tau}
\end{pmatrix}
=
\begin{pmatrix}
\rho\dT{I}&\tTd{K}&0&0 \\
\tTd{K}^{T}&\qT{J}&0&0\\
0&0&\qT{C}&\cTd{M}\\
0&0&\cTd{M}^T&\sT{A}
\end{pmatrix}
\begin{pmatrix}
\V{v}\\
\dT{\nabla v}\\
\dT{\varepsilon}\\
\tTd{\eta}
\end{pmatrix}\label{eq:ConstMatrix}
\een
where
\begin{itemize}
\item $\rho I_{(ij)}$ is the macroscopic mass density;
\item $K_{ijk}$ is the coupling inertia tensor;
\item $J_{ijqr}$ is the second order inertia tensor.
\item $C_{\underline{(ij)}\  \underline{(lm)}}$ is the classical elasticity tensor;
\item $M_{(ij)(lm)n}$ is a fifth-order coupling elasticity tensor;
\item $A_{\underline{(ij)k}\ \underline{(lm)n}}$ a six-order tensor. 
\end{itemize}
As can be observed on the matricial form of the constitutive law (as presented in Eq. \eqref{eq:ConstMatrix}), we implicitly made the hypothesis that space and time are not coupled by the constitutive law.
Albeit being quite unusual, this coupling may appear under certain circumstances. Media in which such phenomena emerge are of Willis type, and more details can be found in \cite{Wil97,NS11,NHA15a}. In the present approach those equations are postulated on phenomenological bases. It should be noted that they can be also derived using a dynamic homogenization procedure \cite{Bacigalupo:2014jt,BG14,NHA15b}

The study of the higher-order elasticity tensors involved in this law has been the object of previous papers \cite{ABB09,ALH13,ADR15}.  
The substitution of the constitutive equations \eqref{eq:Const1} and \eqref{eq:Const2} into the Eq. \eqref{eq:effectivestress} gives
\[
s_{ij}=C_{ijlm}\varepsilon_{lm}+M_{ijklm}^{\sharp}\varepsilon_{lm,k}-A_{ijklmn}\varepsilon_{lm,kn}
 \]
where the dynamic coupling tensor $M_{ijklm}^{\sharp}=M_{ijklm}-M_{klijm}$ has been introduced. In physics this tensor is known as the acoustical gyrotropic tensor and is responsible for the so-called acoustical activity \cite{PB68,Sri88,ADR15}.
Then, using this result into the balance equation \eqref{eq:Bulk} gives
\begin{equation}
C_{ijlm}u_{l,jm}+M_{ijklm}^{\sharp}u_{l,jkm}-A_{ijklmn}u_{l,jkmn}=
\rho \ddot{u}_i+K^{\sharp}_{ipq}\ddot{u}_{p,q}-J_{ipqr}\ddot{u}_{q,pr}
\label{eq:Bulk2}
\end{equation}
where the dynamic coupling inertia tensor $K^{\sharp}_{ipq}=K_{ipq}-K_{piq}$ has been introduced.
It is important to remark that only the dynamic tensors $\cTd{M}^\sharp$ and $\tTd{K}^\sharp$, which have stronger symmetry requirements than their static counterpart, are present in the balance equation. Hence, for some material symmetries $\cTd{M}^\sharp$ and $\tTd{K}^\sharp$ may vanish while $\cTd{M}$ and $\tTd{K}$ are not null tensors \cite{ADR15}. Those odd-order tensors vanish in 3D space for centro-symmetric media, and in 2D space for media that are invariant by a rotation of even-order. To avoid cumbersome expressions this last hypothesis will be assumed in the following. In 2D space, this assumption is not too restrictive since $\cTd{M}$ and $\tTd{K}$ are null in many common situations, and does not preclude chiral behaviors \cite{ADR15}. 

A major result from \cite{ADR15} is that in 2D there are $14$ non equivalent types of anisotropy that can be described by SGE\footnote{The same can be also given for 3D SGE, but the classification would be far more involved, and is not relevant for the present discussion.}. Those different type of anisotropy, together with their number of independent components, are reported  in the following table:
\begin{table}[H]
\begin{center}
\begin{tabular}{|c||c|c|c|c|c|c|c|c|}
\hline
Name & {\scriptsize Oblique} & {\scriptsize Rectangular}& {\scriptsize Digonal} &  {\scriptsize Orthotropic} & {\scriptsize Trichiral} & {\scriptsize Trigonal}  & {\scriptsize Tetrachiral} & {\scriptsize Tetragonal}\\ \hline 
$\lbrack \mathrm{G}_{\mathcal{L}}]$ & $[\Id]$ &$[\ZZ^{\pi}_{2}]$ &$[\ZZ_{2}]$& $[\DD_{2}]$&$[\ZZ_{3}]$ & $[\DD_{3}]$ & $[\ZZ_{4}]$ & $[\DD_{4}]$\\ \hline
$\#_{\mathrm{indep}}(\mathcal{L})$ & $45\ (44)$ & $27 $&$36\ (35)$ & $16$  & $15\ (14)$ & $10$ & $13\ (12)$ & $9$ \\ \hline \hline

Name & {\scriptsize Pentachiral} & {\scriptsize Pentagonal}&{\scriptsize Hexachiral} & {\scriptsize Hexagonal} &  {\scriptsize Hemitropic}& {\scriptsize Isotropic}&&\\ \hline 
$\lbrack \mathrm{G}_{\mathcal{L}}]$  &$[\ZZ_{5}]$ & $[\DD_{5}]$&$[\ZZ_{6}]$ & $[\DD_{6}]$& $[\SO(2)]$& $[\OO(2)]$&&\\ \hline
$\#_{\mathrm{indep}}(\mathcal{L})$ & $9\ (8)$ & $7$ & $9\ (8)$ & $7$& $7$ & $6$&&\\ \hline 
\end{tabular}
\end{center}
\caption{The names, the sets of subgroups $[\mathrm{G}_{\mathcal{L}}]$ and the numbers of independent components $\#_{\mathrm{indep}}(\mathcal{L})$ for the 14 symmetry classes of $\mathcal{L}$, where  $\mathcal{L}$ is the constitutive law. The in-parenthesis number indicates the minimal number of components of the law in an appropriate basis.}
\end{table}
in which
 \begin{itemize}
\item $\ZZ_{n}$, for cyclic groups, means that the object is only invariant by $n$-fold rotations. $\ZZ_{n}$-invariant objects are said to be chiral ;
\item $\DD_{n}$, for dihedral groups indicates a $n$-fold invariant object that possesses also mirrors perpendicular to the rotation axis. $\DD_{n}$-invariant objects are achiral.
\end{itemize}
Hence, as can be read from the table, SGE is
\begin{enumerate}
\item anisotropic for materials that are $6$-fold invariant;
\item sensitive to the chirality of the matter.
\end{enumerate}
In the next sections we will investigate how these specific features influence wave propagation.

\section{Plane wave propagation in an anisotropic strain-gradient continuum} \label{S:PlW}
Objective of this section is to study plane wave propagation in the framework of anisotropic strain-gradient elasticity. To do so, the classic concept of acoustic tensor has to be revisited. This novel generalized acoustic tensor will be used as main analysis tool. Before going into details of the strain-gradient case, it is useful to make some broader considerations about the physical meaning and the interpretation of the different velocities that characterize wave propagation.

\subsection{Wave propagation and wave velocities in anisotropic dispersive media}\label{S:PlaPro}

\begin{figure}
\centering
\includegraphics[scale=1]{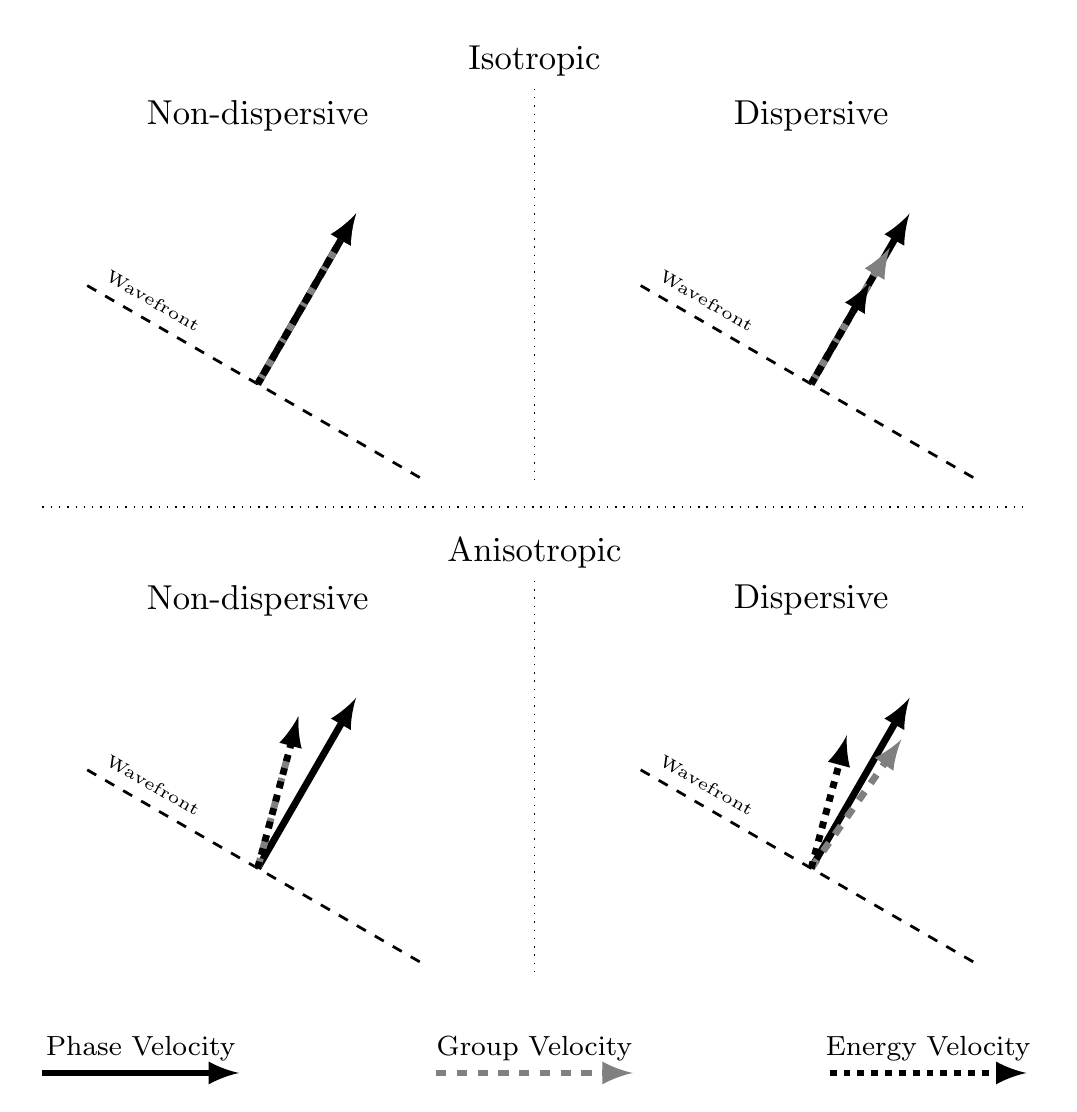}
\caption{Graphical representation of the respective values of phase, energy and group velocity in different cases.}
\label{fig:VelDisp}
\end{figure}

When studying anisotropic materials, useful considerations can be drawn from the analysis of bulk plane waves. This means to seek for solutions of the dynamic equation \eqref{eq:Bulk} in the following form:
\begin{equation}
\V{u}=\V{F}\left(\omega t-\V{k}\cdot\V{x}\right),\label{eq:PWS}
\end{equation} 
where $\V{F}$ is a vector function, $\omega$ the circular frequency, $\V{k}$ the wave vector and $\V{x}$ the position vector.
As well known, plane wave propagation is characterized by different notions of velocity. A priori, these quantities may, or not, be identical. In full generality, four different physical velocities emerge \cite{Bri60}:
\begin{itemize}
\item the phase velocity $\V{v}^{p}$: this quantity is defined as the ratio of the circular frequency $\omega$ over the wave vector $\V{k}$:
\ben
\V{v}^{p}=\frac{\omega}{k}\hat{\V{\xi}}
\een
where $k=\left\|\V{k}\right\|$ is the wave number associated to $\V{k}$, and $\hat{\V{\xi}}$ the unit vector in the direction of $\V{k}$, so that $\V{k}=k\hat{\V{\xi}}$. This is a secant velocity that describes the speed of the wavefront of single harmonic wave oscillations for a wave propagating toward the direction $\hat{\V{\xi}}$.
\item the group velocity $\V{v}^{g}$: this second notion, which is a tangent one, describes, in 1-D, the modulation of the signal:
\ben
\V{v}^{g}=\dfrac{\partial \omega}{\partial \V{k}}=\nabla_{\V{k}}\omega
\een 
This velocity is related to the modulation of the wave packet, and is a kind of "particle" related velocity.
\item the energy velocity $\V{v}^{e}$: this velocity deals with the energy flow within the medium and hence is defined using the Poynting vector. 
\ben
\V{v}^{e}=\dfrac{\V{P}}{\mathcal{E}}
\een
where $\V{P}$ is the Poynting vector while $\mathcal{E}=\mathcal{K}+\mathcal{P}$ is the total energy density, sum of potential and kinetic energy. 
\item the signal velocity $\V{v}^{s}$: this velocity is related to the propagation of the information. This notion, introduced by Sommerfeld in  \cite{Bri60}, is, contrary to the others, bound by relativity principle.
\end{itemize}
In the classical situation of an isotropic, linear, homogeneous, non dispersive, non dissipative medium these four velocities are identical. However, once one of these hypotheses is modified, this equality is not true anymore. In the case of an anisotropic medium (all other hypotheses being conserved), for example, the phase velocity differs from the 3 others, i.e.
\ben
\V{v}^{p}\neq \V{v}^{g}=\V{v}^{s}=\V{v}^{e}
\een
A general summary of relationships between phase, group and energy velocities, in the case of dispersive and non dispersive media, can be found in Table \ref{tab:vel} and illustrated Fig.\ref{fig:VelDisp}. As it can be noticed, in the case of dispersive media, group and energy velocity are not anymore equal to each other. This must lead to a reinterpretation of group velocity with respect to the classic non-dispersive case. At the end of Section \ref{S:MSGE} these properties will be verified analytically and numerically for an elastic strain gradient continuum.
\begin{table}[h]
\centering
\begin{tabular}{|c|c|c|c|}
\hline 
 \multicolumn{2}{|c|}{Isotropic} & \multicolumn{2}{|c|}{Anisotropic}  \\ 
\hline 
 Non dispersive	& Dispersive	& Non dispersive	& Dispersive \\
\hline 
 $\V{v}^{p}=\V{v}^{g}=\V{v}^{e}$ 			& $\V{v}^{p}\neq \V{v}^{g}\neq\V{v}^{e}$ 		&  $\V{v}^{p}\neq \V{v}^{g}=\V{v}^{e}$			&  $\V{v}^{p}\neq \V{v}^{g}\neq \V{v}^{e}$ 	\\ 
\hline 
\end{tabular} 
\caption{Summary of relationships between velocities. }\label{tab:vel}
\end{table}

\subsection{The generalized acoustic tensor}
To obtain the different velocities of a plane wave in the framework of SGE, let us consider the following plane wave solution, equivalent to \eqref{eq:PWS},
\begin{equation}
u_{i}=U_{i} \mathcal{A} \exp\left[\imath\omega\left(t-\dfrac{1}{V}\hat{\xi}_{i}x_{i}\right)\right]\label{eq:WaveSol}
\end{equation}
where $V=\left\|\V{v}^{p}\right\|$ is the norm of the phase velocity of the wave-front and $\V{\hat{\xi}}$ the unit vector pointing toward the direction of propagation, i.e. the normal to the wave-front. Moreover, $U_i$ is a real valued unitary vector representing the polarization (direction of motion) and $\mathcal{A}$ is a complex amplitude. These quantities are both independent of $x_i$ and $t$. Phase velocity and wave-vector can be conveniently summed-up in one quantity, namely the slowness vector $\xi_{i}$: 
\ben
\xi_{i}=\frac{1}{V}\hat{\xi}_{i}.
\een 
The substitution of ansatz \eqref{eq:WaveSol} into the balance equation \eqref{eq:Bulk2} yields
\ben
\left(\left (C_{ijlm} -\omega^{2} J_{ijlm}\right )\hat{\xi}_{j}\hat{\xi}_{m}
+\frac{\omega^{2}}{V^{2}}A_{ijklmn}\hat{\xi}_{j}\hat{\xi}_{k}\hat{\xi}_{m}\hat{\xi}_{n}\right)U_{l}=\rho V^{2} U_{i},
 \een
which can be conveniently rewritten as
\begin{equation}
\hat{Q}_{il}U_{l}=\rho V^{2}U_{i},\label{eq:EVP}
\end{equation}
where the generalized acoustic tensor $\hat{Q}_{il}$ is defined as follows:
\begin{equation}
\hat{Q}_{il}=\left (C_{ijlm} -\omega^{2} J_{ijlm}\right )\hat{\xi}_{j}\hat{\xi}_{m}
+\frac{\omega^{2}}{V^{2}}A_{ijklmn}\hat{\xi}_{j}\hat{\xi}_{k}\hat{\xi}_{m}\hat{\xi}_{n}. \label{eq:BulkPlane}
\end{equation}
As can be noticed, the classic definition of the acoustic tensor is retrieved (i.~e. $Q_{il}=C_{ijlm}\hat{\xi}_{j}\hat{\xi}_{m}$) in the following situations:
\begin{itemize}
\item when the tensor $A_{ijklmn}$  vanishes, that is for a classic continuum;
\item when $\omega\rightarrow 0$, that is for low frequencies.
\end{itemize}
From the solution of the eigenvalue problem associated to Eqn. \eqref{eq:EVP}, it is possible to obtain useful information concerning phase velocity and polarization of plane waves propagating with a wavefront perpendicular to a given direction $\hat{\xi_i}$. Moreover, from the polar plot of the slowness, one can compute the so- called slowness surfaces (or curves in 2D).

\subsection{Conservation of energy, Poynting vector and energy velocity}
As previously discussed, energy velocity is usually considered as being equivalent to group velocity. Since we have pointed out that this interchangeability is no more true for dispersive media, the two notions have to be distinguished. To that aim let us first compute the velocity at which the energy carried by the plane wave is propagating. 

Since we are working with harmonic plane waves, the energy velocity vector $\V{v}^{g}$ is defined by the relation 
\begin{equation}
v^{e}_i=\dfrac{\left <P_i \right >}{\left <\mathcal{E} \right >}\label{eg:EnVelDef}
\end{equation}
where the definition 
\ben
\left <\cdot \right >=\dfrac{1}{T}\int_{0}^{T}\cdot \;\text{d}t
\een
is used to compute the mean values over the $T$-period.  In terms of these two quantities, the local form of the conservation of energy reads:
\ben
\frac{\partial \mathcal{E}}{\partial t}+P_{i,i}=0.
\een
Using the constitutive equations \eqref{eq:Const1} and \eqref{eq:Const2}, and the conservation of linear momentum \eqref{eq:Bulk} the expression of $\V{P}$ can be specified:
\ben
P_{j}=-\left(s_{ij}+\dot{q}_{ij}\right)\dot{u}_{i}-\tau_{ikj}\dot{u}_{i,k}.
\een
Inserting the plane wave solution \eqref{eq:WaveSol} into this last expression and performing a temporal averaging,  the following result is obtained
\ben
\left <P_j \right >=\dfrac{\vert \mathcal{A} \vert ^2}{2}\frac{\omega^{2}}{V}Q_{ijl}^{\flat} U_{l} U_{i}  ,
\een
in which the following tensors have been introduced 
\ben 
Q_{ijl}^{\flat}=\left (C_{ijlm}-\omega^{2}J_{ijlm} \right )\hat{\xi}_{m}-\frac{\omega^{2}}{V^{2}}A^{\flat}_{ijklmn}\hat{\xi}_{k}\hat{\xi}_{m}\hat{\xi}_{n},
\een
with
\ben
A^{\flat}_{ijklmn}=\left(A_{ijklmn}-A_{ikjlmn}\right).
\een
For the kinetic energy and potential energy, the same computation leads to:
\ben
\left <\mathcal{K} \right >=\frac{\vert \mathcal{A} \vert ^2}{4}\omega^{2}\left(\rho\delta_{ik}+\frac{\omega^{2}}{V^{2}}J_{ijkl}\hat{\xi}_{j}\hat{\xi}_{l}\right) U_{k} U_{i}, \qquad \left <\mathcal{P} \right >=\frac{\vert \mathcal{A} \vert ^2}{4}\frac{\omega^{2}}{V^{2}} \tilde{Q}_{il} U_{l} U_{i},
\een  
 where
 \begin{equation}
\tilde{Q}_{il}=C_{ijlm}\hat{\xi}_{j}\hat{\xi}_{m}
+\frac{\omega^{2}}{V^{2}}A_{ijklmn}\hat{\xi}_{j}\hat{\xi}_{k}\hat{\xi}_{m}\hat{\xi}_{n}. 
\end{equation}
Using Eqn. \eqref{eq:EVP}, this last result can be transformed to
\ben
\left <\mathcal{P} \right >=\frac{\vert \mathcal{A} \vert ^2}{4}\omega^{2}\left(\rho\delta_{ik}+\frac{\omega^{2}}{V^{2}}J_{ijkl}\hat{\xi}_{j}\hat{\xi}_{l}\right)U_{k} U_{i}=\left <\mathcal{K} \right >,
\een 
showing that the classical property of the equi-distribution of energy over the period between potential and kinetic energy is verified also into the strain-gradient framework. Then the total energy reads:
\ben
\left <\mathcal{E} \right >=\frac{\vert \mathcal{A} \vert ^2}{2}\omega^{2}\left(\rho\delta_{ik}+\frac{\omega^{2}}{V^{2}}J_{ijkl}\hat{\xi}_{j}\hat{\xi}_{l}\right) U_{k} U_{i} .
\een
Finally, using the definition \eqref{eg:EnVelDef}, the expression for the energy velocity is obtained
\begin{equation}
v^{e}_j=\frac{Q_{ijl}^{\flat}U_{l} U_{i}}
{V\left(\rho\delta_{ik}+\frac{\omega^{2}}{V^{2}}J_{ijkl}\hat{\xi}_{j}\hat{\xi}_{l}\right) U_{k}U_{i}}.\label{eq:EnVel}
\end{equation}
 
\subsection{Group velocity}
In this section we compute the group velocity, which we recall is defined as 
\begin{equation}
	\V{v}^{g}=\dfrac{\partial \omega}{\partial \V{k}}.
\end{equation}
From equation \eqref{eq:BulkPlane} it can be shown that
\begin{equation}
v^{g}_j=\frac{Q_{ijl}^{\sharp} U_{l} U_{i}}
{V\left(\rho\delta_{ik}+\frac{\omega^{2}}{V^{2}}J_{ijkl}\hat{\xi}_{j}\hat{\xi}_{l}\right)  U_{k}U_{i} }\label{eq:GrVel}
\end{equation}
where
\ben 
Q_{ijl}^{\sharp}=\left (C_{ijlm}-\omega^{2}J_{ijlm} \right )\hat{\xi}_{m}+\frac{\omega^{2}}{V^{2}}A^{\sharp}_{ijklmn}\hat{\xi}_{k}\hat{\xi}_{m}\hat{\xi}_{n},
\een
and
\begin{equation}
A^{\sharp}_{ijklmn}=\left(A_{ikjlmn}+A_{ijklmn}\right)
\end{equation}
As expected, the expression of the group velocity, Eq. \eqref{eq:GrVel}, is different from that of the energy velocity(Eq. \eqref{eq:EnVel}).  Moreover, their difference is easily computed, and reads
\be
v_{j}^{g}-v_{j}^{e}=\frac{2\omega^2}{V^3\left(\rho\delta_{ik}+\frac{\omega^{2}}{V^{2}}J_{ijkl}\hat{\xi}_{j}\hat{\xi}_{l}\right) U_{k}U_{i}}A_{ijklmn}U_{l}U_{i}\hat{\xi}_{n}\hat{\xi}_{k}\hat{\xi}_{m}\label{eq:DiffVel}
\ee
This expression confirms that in strain gradient continua the energy and group velocities are not identical and shows that the usual association between group velocity and energy velocity is retrieved in the low frequency limit. 

\subsection{Synthesis}
In this section it has been shown that for a SGE continuum group and energy velocity have to be distinguished. In the case of a centrosymmetric medium, the difference between those two notions is directly related to the second-order elasticity tensor $\sT{A}$. As a consequence, and since SGE is a long wavelength approximation of the elasticity of heterogeneous materials, those two notions should be kept different for architectured materials as soon as micro-structural effects are involved.

\section{Case studies}\label{S:HexMat}

This section is devoted to the analysis of some common situations that have been chosen to illustrate peculiar features of SGE. The following case studies will be analyzed:
\begin{itemize}
\item 2D hexagonal ($\DD_{6}$) lattice
\item 2D hexachiral ($\ZZ_{6}$) lattice
\end{itemize}

\noindent\textbf{Material and physical anisotropy} \\
Hexachiral materials, whose unit cell is represented in Fig. \ref{fig:AuxZ6}, are well known for being auxetic, as they possess a negative Poisson module \cite{PL97,DFJ13}. As can be directly observed, the unit cell is only invariant by $6$-fold rotations, but does not have any line of mirror symmetry. The pattern is then said to be chiral. In the language of group theory the point group of the pattern is conjugate to $\ZZ_{6}$ \cite{ADR15}. If mirrors are added to the set of symmetry elements, the pattern becomes achiral and the classical hexagonal honeycomb tiling is retrieved. The point group is now conjugate to $\DD_{6}$. As already discussed in section \ref{S:MSGE} the SGE behavior is different for these two cases and the associated "shapes" for the elastic operator are\footnote{The notation $\mathrm{T}_{G}$ indicates that the tensor $\mathrm{T}$ is $G$-invariant, where $G$ denotes a subgroup of the full orthogonal group, i.e. $G\subseteq\OO(2)$.} \cite{ADR15}

\ben
\mathcal{L}_{\ZZ_{6}}=%
\begin{pmatrix}
\mathrm{C}_{\OO(2)} & \mathrm{0}\\
\mathrm{0} & \mathrm{A}_{\ZZ_{6}} 
\end{pmatrix}
,\qquad
\mathcal{L}_{\DD_{6}}=%
\begin{pmatrix}
\mathrm{C}_{\OO(2)} & \mathrm{0}\\
\mathrm{0} & \mathrm{A}_{\DD_{6}} 
\end{pmatrix}
\een
In both case the classical elasticity is isotropic, and in any rectangular basis its tensor has the following matricial expression:
\begin{equation}
\mathrm{C}_{\OO(2)}=
\begin{pmatrix}
c_{11}&c_{12} &0    \\

      &c_{11} &0    \\
      
      &       &c_{11}-c_{12}    \\
\end{pmatrix}
\end{equation}
For the second order elasticity tensors their matrix expression in bases adapted with the symmetry elements of the microstructure are\footnote{We refer to \cite{ADR15} for a discussion on that topic, see also appendix \ref{s:OrtOrd} for details concerning the orthogonal basis associated to the matrix representation.}

\begin{equation}
\mathrm{A}_{\ZZ_{6}}=
\begin{pmatrix}
\scriptstyle a_{11}&\scriptstyle a_{12}&\scriptstyle \frac{a_{11}-a_{22}}{\sqrt{2}}-a_{23}&  \scriptstyle 0&\scriptstyle a_{15}             &\scriptstyle-\frac{a_{15}}{\sqrt{2}}\\
                                        
                 &\scriptstyle a_{22}&\scriptstyle a_{23}&\scriptstyle -a_{15}&\scriptstyle 0  &\scriptstyle-\frac{a_{15}}{\sqrt{2}}\\    
                 &               &\scriptstyle \frac{a_{11}+a_{22}}{2}-a_{12}&\scriptstyle\frac{a_{15}}{\sqrt{2}}&\scriptstyle\frac{a_{15}}{\sqrt{2}}&\scriptstyle 0\\
                 &                  &               &\scriptstyle a_{44}& \scriptstyle a_{11}-a_{44}+a_{12} &\scriptstyle \frac{3a_{11}-a_{22}}{\sqrt{2}}-a_{23}-\sqrt{2}a_{44}\\
                    &                  &               &                   & \scriptstyle a_{22}+a_{44}-a_{11} & \scriptstyle \sqrt{2}(a_{44}-a_{11})+a_{23}\\
                    &                  &               &                   &                                   &\scriptstyle \frac{-3a_{11}+a_{22}}{2}-a_{12}+2a_{44}
\end{pmatrix}
\end{equation}

\begin{equation}
A_{\DD_{6}}=
\begin{pmatrix}
\scriptstyle a_{11}&\scriptstyle a_{12}&\scriptstyle \frac{a_{11}-a_{22}}{\sqrt{2}}-a_{23}&  \scriptstyle 0&\scriptstyle 0 &\scriptstyle0\\
                                        
                    &\scriptstyle a_{22}&\scriptstyle a_{23}&\scriptstyle 0&\scriptstyle 0  &\scriptstyle 0\\    
                    &               &\scriptstyle \frac{a_{11}+a_{22}}{2}-a_{12}&\scriptstyle 0&\scriptstyle 0&\scriptstyle 0\\
                    &               &               &\scriptstyle a_{44}& \scriptstyle a_{11}-a_{44}+a_{12} &\scriptstyle \frac{3a_{11}-a_{22}}{\sqrt{2}}-a_{23}-\sqrt{2}a_{44}\\
                    &                  &               &                   & \scriptstyle a_{22}+a_{44}-a_{11} & \scriptstyle \sqrt{2}(a_{44}-a_{11})+a_{23}\\
                    &                  &               &                   &                                   &\scriptstyle \frac{-3a_{11}+a_{22}}{2}-a_{12}+2a_{44}
\end{pmatrix}
\end{equation}
In both cases, the inertia tensor has been considered isotropic and has been replaced by the scalar quantity $\zeta$. In general, different values of micro inertia should be considered for shear and pressure waves. However, given the qualitative nature of the present study, this approximation will not affect our analysis.

\noindent\textbf{Computational procedure} \\
The parameters used in our computation are listed in Table \ref{tab:param}.  They were obtained by performing FE computations on the unit cell using quadratic boundary conditions, and following a procedure described in \cite{ABB10}. These values should be considered as a qualitative approximation of those of the actual material, but they allow us to verify basic properties of the model \cite{FT11,TJA+12}. A forthcoming study will be devoted to a more precise estimation of those parameters. It should be noted that recently some experimental identifications of those coefficient have been conducted using full field measurement and DIC \cite{RKB+15}. 
\\
The computational procedure follows these steps:
\begin{enumerate}
\item the homogenized coefficients of the SGE are computed using FE simulations ;
\item equation \eqref{eq:EVP} is put in the following form
\begin{equation}
\left (\hat{Q}_{il}-\rho V^{2} \delta_{il}\right )U_{l}=0
\end{equation}
and for a given circular frequency $\omega$ we compute the values of the phase velocity $V$ for which the determinant $\det\left (\hat{Q}_{il}-\rho V^{2} \delta_{il}\right )$ vanishes.
\item the nullspace of $\hat{Q}_{il}$ for each couple $\left(\omega, V \right )$ is computed to retrieve the polarization vectors corresponding to each phase velocity at a given frequency. 
\item from the phase velocity the dispersion curves are obtained;
\item phase velocities and polarization vectors are used in Eqs. \eqref{eq:EnVel} and \eqref{eq:GrVel} to obtain energy and group velocities.
\end{enumerate} 

Since we are in a 2D representation, for each value of $\omega$ we find two eigenvalues related to phase velocities of the two allowed wave solutions, along with the associated eigenvectors representing their polarization. At low frequencies, the first solution (mode 1) corresponds to a pure shear ($S-$) mode, and the second solution (mode 2) to a pure pressure ($P-$) mode. For higher frequencies, and depending on the direction of propagation, veering effects \cite{Perkins:1986up} may occur and these modes can be of mixed nature, or even switch. It is important to remark that both velocity and polarization are now function of  both the direction of propagation and the frequency.\\

\begin{figure}
\centering
\includegraphics[width=0.5\linewidth]{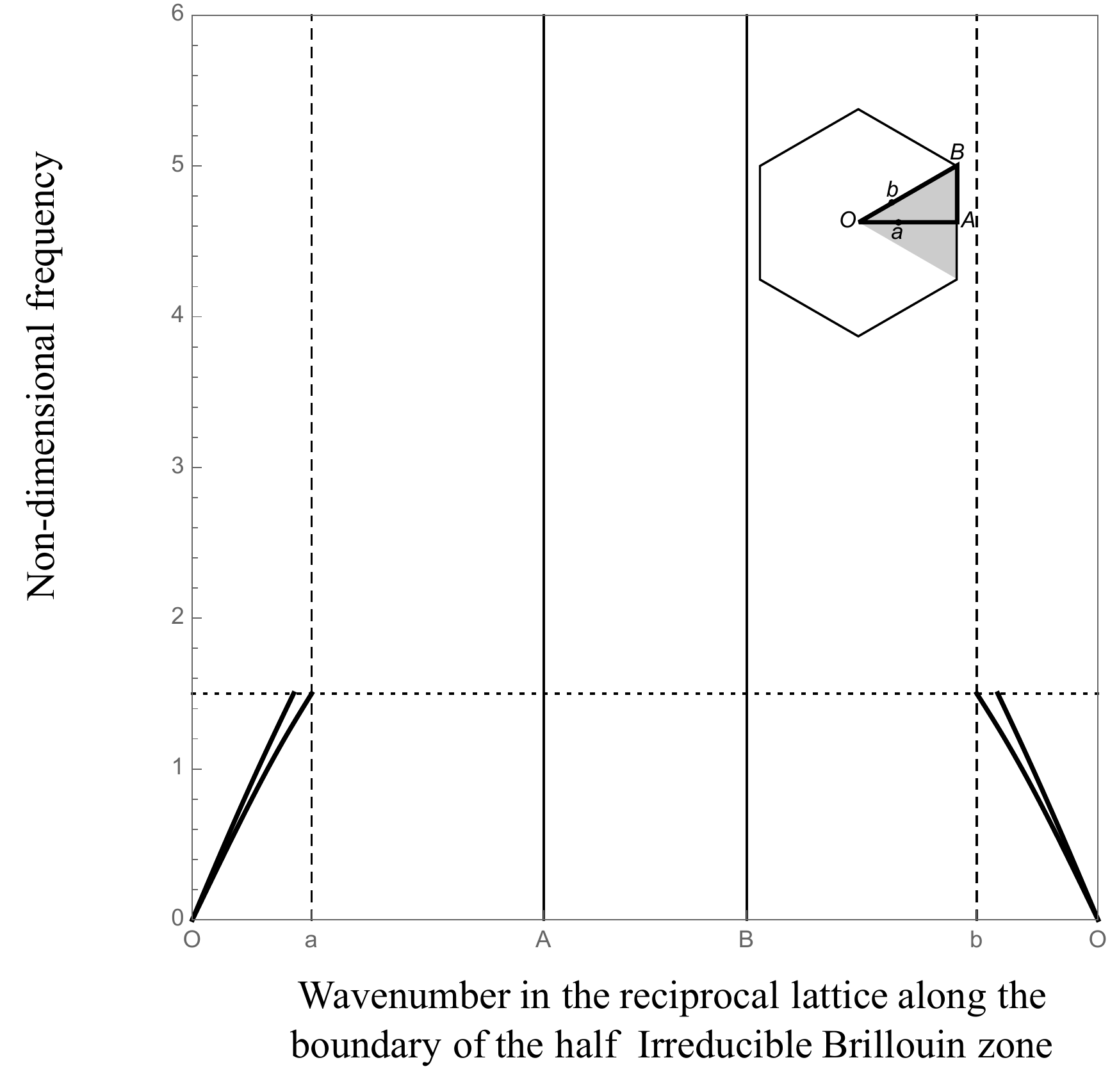}
\caption{Dispersion curves for the hexachiral materials using the strain gradient model, at the edge of the half Irreducible Brillouin zone.}
\label{fig:DispZ6Bri}
\end{figure}

\noindent\textbf{Result analysis}\\
Let us start the analysis of the results from the dispersion curves. In Fig. \ref{fig:DispZ6Bri}, the dispersion curves are computed following the edge of the half Irreducible Brillouin zone, within the limits of validity of the model. As can be seen comparing Fig. \ref{fig:DispZ6Bri} and Fig. \ref{fig:HexDispZones}, the long wave portion of the acoustic branches is qualitatevely well captured. The estimation of the zone of validity for the SGE model is strongly related to that of the parameters, that is why we stress the fact that the results here presented should be considered as qualitative. The circular frequency has been normalized with respect to the resonance frequency $\Omega_0=10.48 \times 10^3\,\text{rad/s}$ of the single ligament of the hexachiral material.

\begin{figure}
\centering
\includegraphics[width=\linewidth]{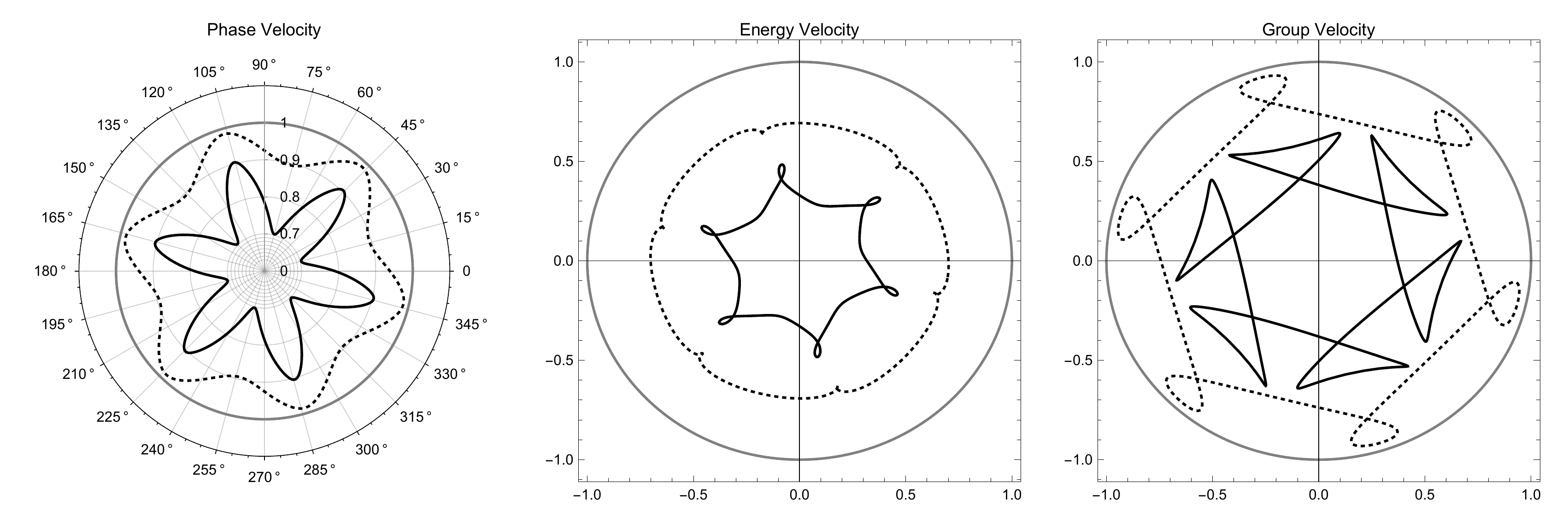}\\
\includegraphics[width=\linewidth]{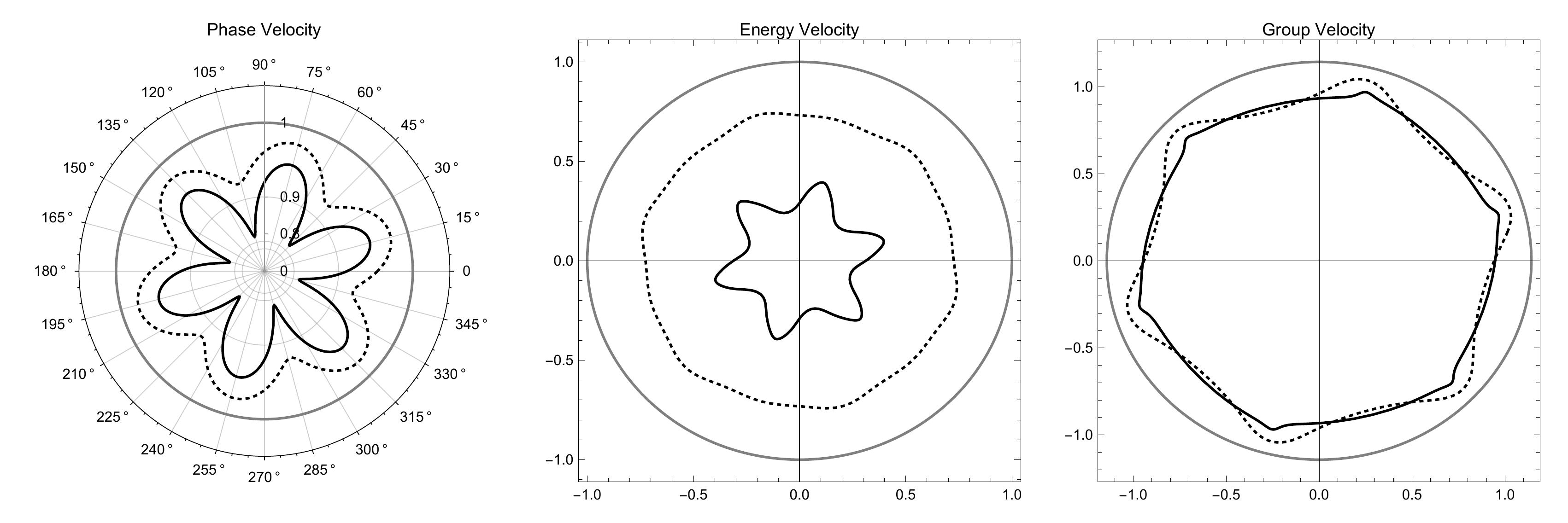}
\caption{Phase, energy and group velocities for a $\ZZ_{6} $ material at $\Omega_1$ (solid gray), $\Omega_2 $ (dashed), $\Omega_3$ (solid black) for the mode 1 (up)  and 2 (down).}
\label{fig:PhEnGrZ6}
\end{figure}

\begin{figure}
\centering
\includegraphics[width=\linewidth]{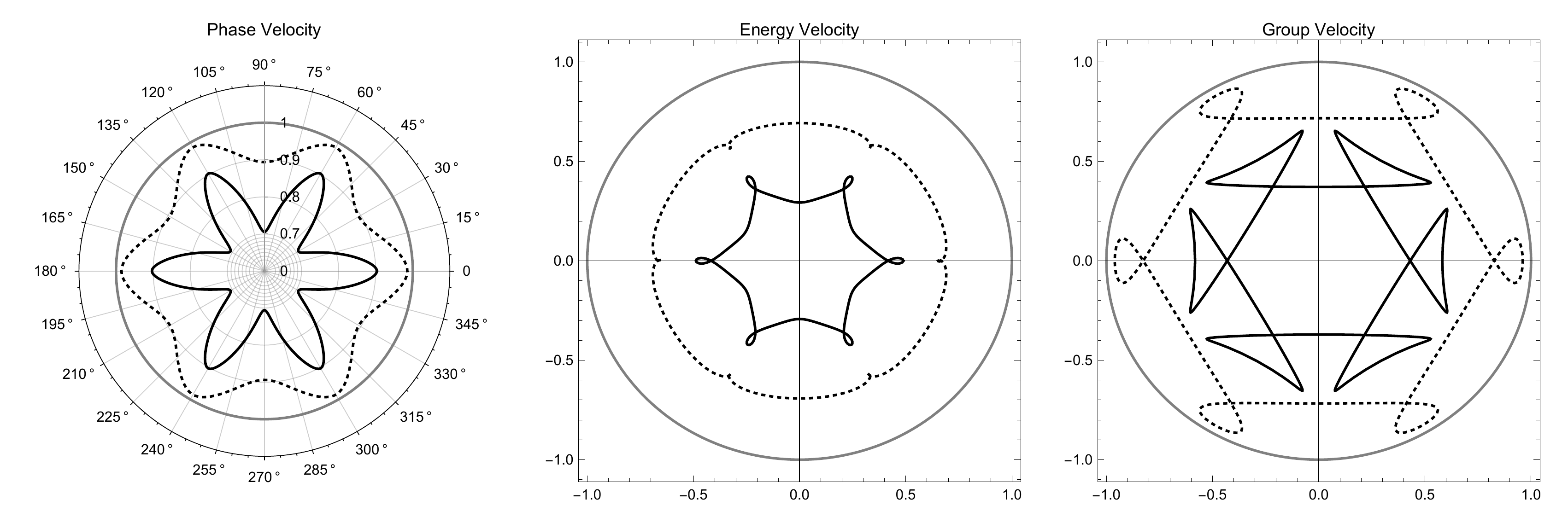}\\
\includegraphics[width=\linewidth]{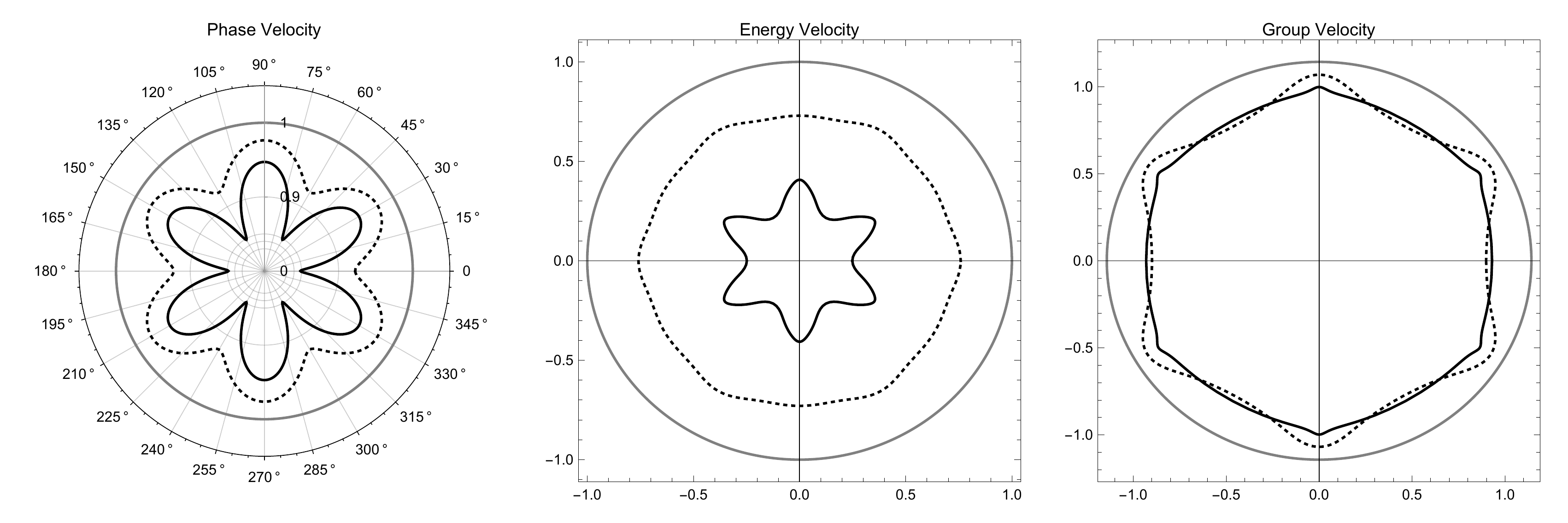}
\caption{Phase, energy and group velocities for a $\DD_{6} $ material at $\Omega_1$ (solid gray), $\Omega_2 $ (dashed), $\Omega_3$ (solid black) for the mode 1 (up)  and 2 (down)..}
\label{fig:PhEnGrD6}
\end{figure}

Useful information can be retrieved from the polar plots of phase, group and energy velocities. These quantities are plotted in Fig.\ref{fig:PhEnGrZ6} for a $\ZZ_{6}$-invariant material and in Fig.\ref{fig:PhEnGrD6} for a $\DD_{6}$-invariant one, for both modes. In these plots, three specific values of frequency have been considered: $\Omega_1=0.01\,\Omega_0$, $\Omega_2=0.8\,\Omega_0 $, $\Omega_3=1.3 \,\Omega_0$. For simplifying the analysis, all velocities have been normalized with respect to the low frequency value of the phase velocity of the first mode. As can be observed, both for hexagonal and hexachiral symmetries, every velocity is isotropic at low frequencies (solid gray lines in Figs. \ref{fig:PhEnGrZ6} and \ref{fig:PhEnGrD6}). Then, when increasing the frequency, a breaking of symmetry occurs, and the specific features of each symmetry class emerge. In particular, from the curves at $\Omega_3$ (solid black lines) one can easily distinguish a hexagonal-like shape. In the case of the $\ZZ_{6}$ class, the chiral effect can also be observed, as each polar curve does not possess any mirror symmetries, while the rotational symmetries are preserved. The chiral effect is not particularly evident due to the qualitative estimation of the coefficients. These are consistent  with those obtained from Bloch analysis, e.~g. in \cite{SRG+09}. Concerning the $\DD_{6}$ class, the plot of the energy velocity is clearly in agreement with the plot presented in Fig \ref{fig:Snapshots-short}. Again, these results are consistent with those from Bloch analysis presented in \cite{CG14}.
It is of major importance to remark that energy and group velocities do not share the same polar plot, thus confirming that they should be treated as two separate quantities. Further studies will be devoted to this distinction, when a better estimation of the coefficients will be available.\\

\begin{table}
\centering
\begin{tabular}{ccc}
 & Hexagonal $\DD_{6} $ &  \\
\hline
Parameter & Value & Unit \\
\hline
$c_{11}$ & $7.3\times 10^7$ & Pa\\ 
$c_{12}$ & $-3.8\times 10^7$ & Pa\\ 
$a_{11}$ & $1.51\times 10^3$ & Pa\\ 
$a_{12}$ & $1.00\times 10^3$ & Pa$\cdot$ m\\ 
$a_{22}$ & $3.52\times 10^3$ & Pa$\cdot$ m\\ 
$a_{15}$ & 0 & Pa$\cdot$ m\\  
$a_{23}$ & $-0.72\times 10^3$ & Pa$\cdot$ m\\ 
$a_{44}$ & $12.96\times 10^3$ & Pa$\cdot$ m\\ 
$\zeta$ & $0.0648$ & kg$\cdot$ m\\ 
$\rho$ & $249.32$ & kg/$\text{m}^3$\\ 
\hline 
\end{tabular} 
\qquad
\begin{tabular}{ccc}
 & Hexachiral $\ZZ_{6}$ &  \\
\hline
Parameter & Value & Unit \\
\hline
$c_{11}$ & $6.58\times 10^7$ & Pa\\ 
$c_{12}$ & $-3.5\times 10^7$ & Pa\\ 
$a_{11}$ & $1.14\times 10^3$ & Pa\\ 
$a_{12}$ & $0.75\times 10^3$ & Pa$\cdot$ m\\ 
$a_{22}$ & $2.64\times 10^3$ & Pa$\cdot$ m\\ 
$a_{15}$ & $-0.84\times 10^3$ & Pa$\cdot$ m\\  
$a_{23}$ & $-0.54\times 10^3$ & Pa$\cdot$ m\\ 
$a_{44}$ & $9.72\times 10^3$ & Pa$\cdot$ m\\ 
$\zeta$ & $0.0648$ & kg$\cdot$ m\\ 
$\rho$ & $249.32$ & kg/$\text{m}^3$\\ 
\hline 
\end{tabular} 
\caption{Values of the parameters used in simulations for the hexagonal $\DD_{6}$ (left) and hexachiral  $\ZZ_{6}$ (right) materials.}
\label{tab:param} 
\end{table}
\section{Conclusions} \label{S:concl}
In the present paper some specific features of wave propagation in SGE media have been studied. This model allows to produce the following effects that can not be modelled classically but which can be experimentally observed and numerically simulated:
\begin{enumerate}
\item Anisotropy of hexagonal lattices;
\item Chiral sensitivity;
\item Distinction between group and energy velocities;
\end{enumerate}
It has to be noted that the first point can not be modelled using a Cosserat (or Micropolar) medium. 

Numerical results concerning free plane wave propagation in hexagonal ($\DD_{6}$) and hexachiral ($\ZZ_{6}$) materials are presented and discussed. These results are consistent with those obtained from Bloch analysis of the unite cell and, concerning the honeycomb, in agreement with experiments. 
\section*{Acknowledgements}
The authors would like to gratefully acknowledge the F\'{e}d\'{e}ration Francilienne de M\'{e}canique for financial support through its starting grant program. Giuseppe Rosi wish to thank the Facult\'{e} de Sciences et Technologie of the Universit\'{e} Paris-Est Cr\'{e}teil Val de Marne for the financial support.
\begin{appendix}

\section{Orthonormal basis and matrix component ordering}\label{s:OrtOrd}

Let be defined the following spaces:
\begin{eqnarray*} \label{3DT}
\mathbb{T}_{(ij)}&=&\{\mathrm{T}\in\mathbb{T}_{ij}|\mathrm{T}=\displaystyle\sum\limits_{i,j=1}^{2 }T_{ij}e_{i}\otimes e_{j},\ T_{ij}=T_{ji}\}\\ 
\mathbb{T}_{(ij)k}&=&\{\mathrm{T}\in\mathbb{T}_{ijk}|\mathrm{T}=\displaystyle\sum\limits_{i,j,k=1}^{2}T_{ijk}e_{i}\otimes e_{j}\otimes e_{k},\ T_{ijk}=T_{jik}\} 
\end{eqnarray*}%
which are, in 2D, respectively,  3- and 6-dimensional vector spaces. Therefore
\begin{itemize}
\item  the first-order elasticity tensor $\qT{C}$ is a self-adjoint endomorphism of $\mathbb{T}_{(ij)}$;
\item  the second-order elasticity tensor $\qT{A}$ is a self-adjoint endomorphism of $\mathbb{T}_{(ij)k}$.
\end{itemize}
In order to express the Cauchy-stress tensor $\dT{\sigma}$, the strain tensor $\dT{\varepsilon}$, the strain-gradient tensor $\qT{\eta}$ and the hyperstress tensor $\tTd{\tau}$ as 3- and 6-dimensional vectors and write $\qT{C}$ and $\sT{A}$ as, respectively: a $3\times 3$ and $6\times 6$ matrices, we introduce the following orthonormal basis vectors:
\begin{eqnarray*}\label{Basis}
\widetilde{e}_{I}&=&\left( \frac{1-\delta _{ij}}{\sqrt{2}}+\frac{\delta _{ij}}{2}\right) \left(e_{i}\otimes e_{j}+e_{j}\otimes e_{i}\right) ,\quad 1\leq I\leq 3\\
\widehat{e}_{\alpha }&=&\left( \frac{1-\delta _{ij}}{\sqrt{2}}+\frac{\delta _{ij}}{2}\right) \left( e_{i}\otimes e_{j}+e_{j}\otimes e_{i}\right) \otimes e_{k},\quad 1\leq \alpha\leq 6  
\end{eqnarray*}%
where the summation convention for a repeated subscript does not apply.
Then, the aforementioned tensors can be expressed as:
\begin{equation}
\widetilde{\varepsilon} =\displaystyle\sum\limits_{I=1}^{3}\widetilde{\varepsilon}_{I}\widetilde{e}_{I},
\quad 
\widetilde{\sigma} = \displaystyle\sum\limits_{I=1}^{3}\widetilde{\sigma}_{I}\widetilde{e}_{I}, \quad  \widehat{\eta} = \displaystyle\sum\limits_{\alpha=1}^{6}\widehat{\eta}_{\alpha}\widehat{e}_{\alpha}, \quad  \widehat{\tau} = \displaystyle\sum\limits_{\alpha=1}^{6}\widehat{\tau}_{\alpha}\widehat{e}_{\alpha}
\end{equation}%
\begin{equation} 
\mathrm{\widetilde{C}}=\displaystyle\sum\limits_{I,J=1,1}^{3,3} \widetilde{C}_{IJ}\widetilde{e}_{I}\otimes \widetilde{\widehat{e}}_{J}   \quad \mathrm{\overline{M}}=\displaystyle\sum\limits_{I,\alpha=1,1}^{3,6} \overline{M}_{I \alpha}\mathbf{\widetilde{e}}_{I}\otimes \mathbf{\widehat{e}}_{\alpha }, \quad
\mathrm{\widehat{A}}=\displaystyle\sum\limits_{\alpha,\beta=1,1}^{6,6} \widehat{A}_{\alpha\beta}\widehat{e}_{\alpha}\otimes \widehat{e}_{\beta },    
\end{equation}%
so that the constitutive law can be written in the matrix form%
\begin{equation}
\begin{cases}
\widetilde{\sigma}_{I}=\widetilde{C}_{IJ}\widetilde{\varepsilon}_{J}\\
\widehat{\tau}_{\alpha}=\widetilde{A}_{\alpha\beta}\widehat{\eta}_{\beta}
\end{cases}
\label{SGER}
\end{equation}%
The relationship between the matrix components $\widetilde{\varepsilon}_{I}$ and $\varepsilon _{ij}$, and between $\widehat{\eta}_{\alpha}$ and $\eta_{ijk}$ are
\begin{equation}
\widetilde{\varepsilon}_{I}=
\begin{cases}
\varepsilon _{ij}\text{ \ if \ }i=j, \\ 
\sqrt{2}\varepsilon _{ij}\text{ \ if \ }i\neq j;%
\end{cases}%
\quad 
\widehat{\eta}_{\alpha }=%
\begin{cases}
\eta_{ijk}\text{ \ if \ }i=j, \\ 
\sqrt{2}\eta_{ijk}\text{ \ if \ }i\neq j;%
\end{cases}%
\label{3-to-1}
\end{equation}
and, obviously, the same relations between $\widetilde{\sigma}_{I}$ and $\sigma _{ij}$ and $\widehat{\tau}_{\alpha}$ and $\tau_{ijk}$ hold. For the constitutive tensors we have the following correspondences: 
\begin{equation}
\widetilde{C}_{IJ}=%
\begin{cases}
C_{ijkl}\text{ \ if \ }i=j\ \text{and}\ k=l\text{,} \\ 
\sqrt{2}C_{ijkl}\text{ \ if \ }i\neq j\ \text{and}\ k=l\ \text{or}\ i=j\ 
\text{and}\ k\neq l, \\ 
2C_{ijkl}\text{ \ if \ }i\neq j\ \text{and}\ k\neq l.%
\end{cases}
\end{equation}%
\begin{equation}
\widehat{A}_{ \alpha\beta }=%
\begin{cases}
A_{ijklmn}\text{ \ if \ }i=j\ \text{and}\ l=m\text{,} \\ 
\sqrt{2}A_{ijklmn}\text{ \ if \ }i\neq j\ \text{and}\ l=m\ \text{or}\ i=j\ 
\text{and}\ l\neq m, \\ 
2A_{ijklmn}\text{ \ if \ }i\neq j\ \text{and}\ l\neq m.
\end{cases}
\label{6-to-3}
\end{equation}
It remains to choose  appropriate two-to-one and three-to-one subscript correspondences between $ij$ and $I$, on one hand, and $ijk$ and $\alpha$, on the other hand. 
For the classical variables the standard two-to-one subscript correspondence is used, i.e:
\begin{table}[H]
\begin{center}
\begin{tabular}{|c||c|c|c|}
\hline
$I$& $1$& $2$& $3$ \\ \hline\hline
$ij$&$11$&$22 $&$12$\\
\hline
\end{tabular}
\end{center}
\caption{The two-to-one subscript correspondence for 2D strain/stress tensors}
\end{table}
with the  following three-to-one subscript correspondence for strain-gradient/hyperstress tensor:
\begin{table}[H]
\begin{center}
\begin{tabular}{|c||c|c|c||c|}
\hline
$\alpha$& $1$& $2$& $3$&  \\ \hline\hline
$ijk $&$111 $&$221 $&$122$& Privileged direction: $1$\\ 
\hline\hline
$\alpha $&$6$&$7$&$8$&  \\ \hline
$ijk $&$222$&$112$&$121 $& Privileged direction: $2$\\ 
\hline\end{tabular}
\end{center}
\caption{The three-to-one subscript correspondence for 2D strain-gradient/hyperstress tensors}
\end{table}

\end{appendix}
\noindent \textbf{References}

\bibliographystyle{apalike_p}
\bibliography{Bib}

\begin{thebibliography}{}

\bibitem[Auffray et~al., 2009]{ABB09}
{ Auffray N., Bouchet R., and Br\'{e}chet Y.} (2009).
\newblock {\em Derivation of anisotropic matrix for bi-dimensional
  strain-gradient elasticity behavior}.
\newblock International Journal of Solids and Structures, vol. 46
  n$^{\circ}\,$2, pp 440--454.

\bibitem[Auffray et~al., 2010]{ABB10}
{ Auffray N., Bouchet R., and Br\'{e}chet Y.} (2010).
\newblock {\em Strain gradient elastic homogenization of bidimensional cellular
  media}.
\newblock International Journal of Solids and Structures, vol. 47
  n$^{\circ}\,$13, pp 1698--1710.

\bibitem[Auffray et~al., 2015a]{AIE+13}
{ Auffray N., dell'Isola F., Eremeyev V.A., Madeo A., and Rosi G.} (2015a).
\newblock {\em Analytical continuum mechanics {\`a} la Hamilton--Piola least
  action principle for second gradient continua and capillary fluids}.
\newblock Mathematics and Mechanics of Solids, vol. 20 n$^{\circ}\,$4, pp
  375--417.

\bibitem[Auffray et~al., 2015b]{ADR15}
{ Auffray N., Dirrenberger J., and Rosi G.} (2015b).
\newblock {\em A complete description of bi-dimensional anisotropic
  strain-gradient elasticity}.
\newblock International Journal of Solids and Structures,  n$^{\circ}\,$69-70,
  pp 195--206.

\bibitem[Auffray et~al., 2013]{ALH13}
{ Auffray N., Le~Quang H., and He Q.C.} (2013).
\newblock {\em Matrix representations for 3D strain-gradient elasticity}.
\newblock Journal of the Mechanics and Physics of Solids, vol. 61
  n$^{\circ}\,$5, pp 1202--1223.

\bibitem[Bacigalupo and Gambarotta, 2014a]{BG14}
{ Bacigalupo A. and Gambarotta L.} (2014a).
\newblock {\em Homogenization of periodic hexa- and tetrachiral cellular
  solids}.
\newblock Composite Structures, vol. 116, pp 461--476.

\bibitem[Bacigalupo and Gambarotta, 2014b]{Bacigalupo:2014jt}
{ Bacigalupo Andrea and Gambarotta Luigi} (2014b).
\newblock {\em {Second-gradient homogenized model for wave propagation in
  heterogeneous periodic media}}.
\newblock International Journal of Solids and Structures, vol. 51
  n$^{\circ}\,$5, pp 1052--1065.

\bibitem[Brillouin, 1960]{Bri60}
{ Brillouin L.} (1960).
\newblock {\em {Wave Propagation and Group Velocity}}, vol. 191.
\newblock Academic Press.

\bibitem[Celli and Gonella, 2014]{CG14}
{ Celli P. and Gonella S.} (2014).
\newblock {\em Laser-enabled experimental wavefield reconstruction in
  two-dimensional phononic crystals}.
\newblock Journal of Sound and Vibration, vol. 333 n$^{\circ}\,$1, pp 114--123.

\bibitem[Cosserat and Cosserat, 1909]{Cos09}
{ Cosserat E. and Cosserat F.} (1909).
\newblock {\em Théorie des corps déformables}.
\newblock Paris.

\bibitem[dell'Isola et~al., 2011]{dellIsola:2011ti}
{ dell'Isola F., Madeo A., and Placidi L.} (2011).
\newblock {\em {Linear plane wave propagation and normal transmission and
  reflection at discontinuity surfaces in second gradient 3D continua}}.
\newblock ZAMM {\textendash} Journal of Applied Mathematics and Mechanics /
  Zeitschrift f{\"u}r Angewandte Mathematik und Mechanik, vol. 92
  n$^{\circ}\,$1, pp 52--71.

\bibitem[Dirrenberger et~al., 2013]{DFJ13}
{ Dirrenberger J., Forest S., and Jeulin D.} (2013).
\newblock {\em Effective elastic properties of auxetic microstructures:
  anisotropy and structural applications}.
\newblock International Journal of Mechanics and Materials in Design, vol. 9
  n$^{\circ}\,$1, pp 21--33.

\bibitem[DiVincenzo, 1986]{DiV86}
{ DiVincenzo D.} (1986).
\newblock {\em {Dispersive corrections to continuum elastic theory in cubic
  crystals.}}
\newblock Physical review. B, Condensed matter, vol. 34 n$^{\circ}\,$8, pp
  5450--5465.

\bibitem[Dresselhaus et~al., 2007]{DDJ08}
{ Dresselhaus M.-S., Dresselhaus G., and Jorio A.} (2007).
\newblock {\em Group theory: application to the physics of condensed matter}.
\newblock Springer Science \& Business Media.

\bibitem[Erigen, 1967]{Eri68}
{ Erigen A.C.} (1967).
\newblock {\em Theory of micropolar elasticity}.
\newblock In~: Fracture, vol. 2.,  ed. Leibowitz H., pp 621--629. Academic
  Press, New York.

\bibitem[Forest and Trinh, 2011]{FT11}
{ Forest S. and Trinh D.K.} (2011).
\newblock {\em {Generalized continua and non-homogeneous boundary conditions in
  homogenisation methods}}.
\newblock ZAMM - Journal of Applied Mathematics and Mechanics / Zeitschrift
  f\"{u}r Angewandte Mathematik und Mechanik, vol. 91 n$^{\circ}\,$2, pp
  90--109.

\bibitem[Gazalet et~al., 2013]{GDK+13}
{ Gazalet J., Dupont S., Kastelik J.-C., Rolland Q., and Djafari-Rouhani B.}
  (2013).
\newblock {\em A tutorial survey on waves propagating in periodic media:
  Electronic, photonic and phononic crystals. Perception of the Bloch theorem
  in both real and Fourier domains}.
\newblock Wave Motion, vol. 50 n$^{\circ}\,$3, pp 619--654.

\bibitem[Germain, 1973]{Ger73}
{ Germain P.} (1973).
\newblock {\em The method of virtual power in continuum mechanics. Part 2:
  Microstructure}.
\newblock SIAM Journal on Applied Mathematics, vol. 25 n$^{\circ}\,$3, pp
  556--575.

\bibitem[Ghiba et~al., 2015]{Ghiba:2014fo}
{ Ghiba I~D, Neff Patrizio, Madeo Angela, Placidi Luca, and Rosi Giuseppe}
  (2015).
\newblock {\em {The relaxed linear micromorphic continuum: Existence,
  uniqueness and continuous dependence in dynamics}}.
\newblock Mathematics and Mechanics of Solids, vol. 20 n$^{\circ}\,$10, pp
  1171--1197.

\bibitem[Gourgiotis et~al., 2013]{Gourgiotis:2013gt}
{ Gourgiotis P.A., Georgiadis H.G., and Neocleous I.} (2013).
\newblock {\em {On the reflection of waves in half-spaces of microstructured
  materials governed by dipolar gradient elasticity}}.
\newblock Wave Motion, vol. 50 n$^{\circ}\,$3, pp 437--455.

\bibitem[Green and Rivlin, 1964]{GR64}
{ Green A.E. and Rivlin R.S.} (1964).
\newblock {\em Multipolar continuum mechanics}.
\newblock Archive for Rational Mechanics and Analysis, vol. 17 n$^{\circ}\,$2,
  pp 113--147.

\bibitem[Liu et~al., 2011]{LHS+11}
{ Liu X.N., Hu G.K., Sun C.T., and Huang G.L.} (2011).
\newblock {\em {Wave propagation characterization and design of two-dimensional
  elastic chiral metacomposite}}.
\newblock Journal of Sound and Vibration, vol. 330 n$^{\circ}\,$11, pp
  2536--2553.

\bibitem[Liu et~al., 2012]{LHH12}
{ Liu X.N., Huang G.L., and Hu G.K.} (2012).
\newblock {\em Chiral effect in plane isotropic micropolar elasticity and its
  application to chiral lattices}.
\newblock Journal of the Mechanics and Physics of Solids, vol. 60
  n$^{\circ}\,$11, pp 1907--1921.

\bibitem[Liu et~al., 2015]{LSS15}
{ Liu Y., Su X., and Sun C.T.} (2015).
\newblock {\em {Broadband elastic metamaterial with single negativity by
  mimicking lattice systems}}.
\newblock Journal of the Mechanics and Physics of Solids, vol. 74, pp 158--174.

\bibitem[Madeo et~al., 2015a]{Madeo:2013uq}
{ Madeo A., Neff P., Ghiba I.D., Placidi L., and Rosi G.} (2015a).
\newblock {\em {Wave propagation in relaxed micromorphic continua: modeling
  metamaterials with frequency band-gaps}}.
\newblock Continuum Mechanics and Thermodynamics, vol. 27, pp 551--570.

\bibitem[Madeo et~al., 2015b]{Madeo:2014df}
{ Madeo A, Neff P, Ghiba I~D, Placidi L, and Rosi G} (2015b).
\newblock {\em {Band gaps in the relaxed linear micromorphic continuum}}.
\newblock ZAMM - Journal of Applied Mathematics and Mechanics / Zeitschrift
  f{\"u}r Angewandte Mathematik und Mechanik, vol. 95 n$^{\circ}\,$9, pp
  880--887.

\bibitem[Madeo et~al., 2014]{Madeo:2014ge}
{ Madeo A, Placidi L, and Rosi Giuseppe} (2014).
\newblock {\em Towards the Design of Metamaterials with Enhanced Damage
  Sensitivity: Second Gradient Porous Materials}.
\newblock Research in Nondestructive Evaluation, vol. 25 n$^{\circ}\,$2, pp
  99--124.

\bibitem[Mindlin, 1964]{Min64}
{ Mindlin R.D.} (1964).
\newblock {\em Micro-structure in linear elasticity}.
\newblock Archive for Rational Mechanics and Analysis, vol. 16 n$^{\circ}\,$1,
  pp 51--78.

\bibitem[Mindlin, 1965]{Min65}
{ Mindlin R.D.} (1965).
\newblock {\em Second gradient of strain and surface-tension in linear
  elasticity}.
\newblock International Journal of Solids and Structures, vol. 1
  n$^{\circ}\,$4, pp 417--438.

\bibitem[Mindlin and Eshel, 1968]{ME68}
{ Mindlin R.D. and Eshel N.N.} (1968).
\newblock {\em On first strain-gradient theories in linear elasticity}.
\newblock International Journal of Solids and Structures, vol. 4
  n$^{\circ}\,$1, pp 109--124.

\bibitem[Nassar et~al., 2015a]{NHA15b}
{ Nassar H., He Q.-C., and Auffray N.} (2015a).
\newblock {\em {On asymptotic elastodynamic homogenization approaches for
  periodic media}}.
\newblock Journal of Mechanics and Physics of Solids.

\bibitem[Nassar et~al., 2015b]{NHA15a}
{ Nassar H., He Q.-C., and Auffray N.} (2015b).
\newblock {\em {Willis elastodynamic homogenization theory revisited for
  periodic media}}.
\newblock Journal of the Mechanics and Physics of Solids, vol. 77, pp 158--178.

\bibitem[Neff et~al., 2014]{Neff:2014ek}
{ Neff P., Ghiba I.D., Madeo A., Placidi L., and Rosi G.} (2014).
\newblock {\em {A unifying perspective: the relaxed linear micromorphic
  continuum}}.
\newblock Continuum Mechanics and Thermodynamics, vol. 26 n$^{\circ}\,$5, pp
  639--681.

\bibitem[Norris and Shuvalov, 2011]{NS11}
{ Norris A.N. and Shuvalov A.L.} (2011).
\newblock {\em {Elastic cloaking theory}}.
\newblock Wave Motion, vol. 48 n$^{\circ}\,$6, pp 525--538.

\bibitem[Perkins and Mote~Jr, 1986]{Perkins:1986up}
{ Perkins N.C. and Mote~Jr C.D.} (1986).
\newblock {\em {Comments on curve veering in eigenvalue problems}}.
\newblock Journal of Sound and Vibration, vol. 106 n$^{\circ}\,$3, pp 451--463.

\bibitem[Phani et~al., 2006]{PWF06}
{ Phani A.S., Woodhouse J., and Fleck N.A.} (2006).
\newblock {\em {Wave propagation in two-dimensional periodic lattices}}.
\newblock The Journal of the Acoustical Society of America, vol. 119
  n$^{\circ}\,$4, pp 1995.

\bibitem[Portigal and Burstein, 1968]{PB68}
{ Portigal D.~L. and Burstein E.} (1968).
\newblock {\em Acoustical activity and other first-order spatial dispersion
  effects in crystals}.
\newblock Physical Review, vol. 170 n$^{\circ}\,$3, pp 673.

\bibitem[Prall and Lakes, 1997]{PL97}
{ Prall D. and Lakes R.S.} (1997).
\newblock {\em Properties of a chiral honeycomb with a poisson's ratio of -1}.
\newblock International Journal of Mechanical Sciences, vol. 39 n$^{\circ}\,$3,
  pp 305--314.

\bibitem[R\'{e}thor\'{e} et~al., 2015]{RKB+15}
{ R\'{e}thor\'{e} J., Kaltenbrunner C., Dand T.C.T., Chaudet P., and Kuhn M.}
  (2015).
\newblock {\em {Gradient-elasticity for honeycomb materials : validation and
  identification from full-field measurements}}.
\newblock International Journal of Solids and Structures, vol. -
  n$^{\circ}\,$72, pp 108--117.

\bibitem[Rosi et~al., 2014]{Rosi:2014il}
{ Rosi G., Nguyen V.-H., and Naili S.} (2014).
\newblock {\em {Reflection of acoustic wave at the interface of a fluid-loaded
  dipolargradient elastic half-space}}.
\newblock Mechanics Research Communications, vol. 56, pp 98--103.

\bibitem[Rosi et~al., 2015]{Rosi:2014hk}
{ Rosi G., Nguyen V.-H., and Naili S.} (2015).
\newblock {\em {Surface waves at the interface between an inviscid fluid and a
  dipolar gradient solid}}.
\newblock Wave Motion, vol. 53 n$^{\circ}\,$0, pp 51--65.

\bibitem[Ruzzene et~al., 2003]{RSS03}
{ Ruzzene M., Scarpa F., and Soranna F.} (2003).
\newblock {\em Wave beaming effects in two-dimensional cellular structures}.
\newblock Smart materials and structures, vol. 12 n$^{\circ}\,$3, pp 363.

\bibitem[Schurig et~al., 2006]{SMJ+06}
{ Schurig D., Mock J.J., Justice B.J., Cummer S.A., Pendry J.B., Starr A.F.,
  and Smith D.R.} (2006).
\newblock {\em Metamaterial electromagnetic cloak at microwave frequencies}.
\newblock Science, vol. 314 n$^{\circ}\,$5801, pp 977--980.

\bibitem[Spadoni et~al., 2009]{SRG+09}
{ Spadoni A., Ruzzene M., Gonella S., and Scarpa F.} (2009).
\newblock {\em Phononic properties of hexagonal chiral lattices}.
\newblock Wave Motion, vol. 46 n$^{\circ}\,$7, pp 435--450.

\bibitem[Srinivasan, 1988]{Sri88}
{ Srinivasan T.P.} (1988).
\newblock {\em A description of acoustical activity using irreducible tensors}.
\newblock Journal of Physics C: Solid State Physics, vol. 21, pp 4207--4219.

\bibitem[Toupin, 1962]{Tou62}
{ Toupin R.A.} (1962).
\newblock {\em Elastic materials with couple-stresses}.
\newblock Archive for Rational Mechanics and Analysis, vol. 11 n$^{\circ}\,$1,
  pp 385--414.

\bibitem[Trinh et~al., 2012]{TJA+12}
{ Trinh D.K., J\"{a}nicke R., Auffray N., Diebels S., and Forest S.} (2012).
\newblock {\em Evaluation of generalized continuum substitution models for
  heterogeneous materials}.
\newblock International Journal for Multiscale Computational Engineering, vol.
  10 n$^{\circ}\,$6.

\bibitem[Willis, 1985]{Wil85}
{ Willis J.R.} (1985).
\newblock {\em {The nonlocal influence of density variations in a composite}}.
\newblock International Journal of Solids and Structures, vol. 21
  n$^{\circ}\,$7, pp 805--817.

\bibitem[Willis, 1997]{Wil97}
{ Willis J.R.} (1997).
\newblock {\em {Dynamics of composites}}.
\newblock In~: Continuum Micromechanics,  ed. Suquet P., chapter Dynamics o, pp
  265--290. Springer-Verlag, New York.

\bibitem[Wolfe, 2005]{Wol05}
{ Wolfe J.P.} (2005).
\newblock {\em Imaging phonons: acoustic wave propagation in solids}.
\newblock Cambridge University Press.

\bibitem[Zachary and Torquato, 2009]{ZT09}
{ Zachary C.E. and Torquato S.} (2009).
\newblock {\em Hyperuniformity in point patterns and two-phase random
  heterogeneous media}.
\newblock Journal of Statistical Mechanics: Theory and Experiment, vol. 2009
  n$^{\circ}\,$12, pp P12015.

\end{thebibliography}

\end{document}